\documentclass[showpacs,showkeys,preprintnumbers,nofootinbib]{revtex4}%
\usepackage{amsfonts}
\usepackage{amsmath}
\usepackage{graphicx}
\usepackage{float}
\usepackage{dcolumn}
\usepackage{bm}
\usepackage{amssymb}
\usepackage{xcolor}%
\providecommand{\U}[1]{\protect\rule{.1in}{.1in}}

\begin{document}
\title{Hybrid phenomenology in a chiral approach}
\author{Walaa I. Eshraim$^{\text{(a,b)}}$, Christian S. Fischer$^{\text{(a,c)}}$,
Francesco Giacosa$^{\text{(d,b)}}$, Denis Parganlija$^{\text{(e)}}$}
\affiliation{$^{\text{(a)}}$ Institute for Theoretical Physics, Justus-Liebig University,
Heinrich-Buff-Ring 16, 35392 Giessen, Germany,}
\affiliation{$^{\text{(b)}}$ Institute for Theoretical Physics, Goethe University,
Max-von-Laue-Str.\ 1, 60438 Frankfurt am Main, Germany,}
\affiliation{$^{\text{(c)}}$ Helmholtz Research Academy Hesse for FAIR, Giessen, Germany}
\affiliation{$^{\text{(d)}}$Institute of Physics, Jan Kochanowski University, ul.
Uniwersytecka 7, 25-406 Kielce, Poland,}
\affiliation{$^{\text{(e)}}$ Institute for Theoretical Physics, Vienna University of
Technology, Wiedner Hauptstr. 8-10, 1040 Vienna, Austria}

\begin{abstract}
We calculate masses and decays of the (lightest) hybrid nonet with exotic
quantum numbers $J^{PC}=1^{-+}$ and the nonet of their chiral partners with
$J^{PC}=1^{+-}$ in the framework of the extended Linear Sigma Model (eLSM). As
an input, we identify $\pi_{1}^{hyb}=\pi_{1}(1600)$ as a low-lying hybrid. We
investigated interaction terms which fulfil chiral symmetry. For what concerns
$\pi_{1}^{hyb},$ the most important decays are $\pi_{1}^{hyb}\rightarrow
b_{1}\pi,$ $\pi_{1}(1600)\rightarrow\rho\pi\eta,$ $\pi_{1}^{hyb}%
\rightarrow\rho\pi,$ and $\pi_{1}^{hyb}\rightarrow KK^{\ast}(892).$ The decays
$\pi_{1}^{hyb}\rightarrow\eta\pi$ and $\pi_{1}^{hyb}\rightarrow\eta^{\prime
}\pi$ are expected to be small but non-zero: they follow from a chirally
symmetric interaction term that breaks explicitly the axial anomaly. For all
the other members of the two hybrid nonets (for which no experimental
candidates exist yet) we report decay ratios that may guide ongoing and future experiments.

\end{abstract}

\maketitle



\section{Introduction}

The search for hybrids is an important part of experimental as well as
theoretical hadronic physics, see e.g.
Refs.~\cite{Meyer:2015eta,Lebed:2016hpi} for reviews. Lattice QCD predicts a
rich spectrum of hybrids below 5 GeV
\cite{Michael:1999ge,Dudek:2009qf,Dudek:2010wm,Dudek:2013yja,Liu:2012ze,Moir:2013ub}%
, but up to now no predominantly hybrid state could be unambiguously assigned
to one of the mesons listed in the PDG \cite{Tanabashi:2018oca}. Yet, two
states with \textquotedblleft exotic\textquotedblright\footnote{Here and in
the following we use the term `exotic' to indicate quantum numbers that are
not possible for quark-anti-quark states in the non-relativistic quark model.
The term \textquotedblleft crypto-exotic\textquotedblright\ is reserved for
mesons with non-exotic quantum numbers, but valence content beyond the
non-relativistic quark model, such as hybrids, gluebalss, and tetraquarks.}
quantum numbers $J^{PC}=1^{-+}$ are listed below 2 GeV: $\pi_{1}(1400)$ and
$\pi_{1}(1600)$. Recent results by COMPASS confirmed the state $\pi_{1}(1600)$
and led to a revival of interest in this topic \cite{Akhunzyanov:2018lqa}. At
the Jefferson Lab (JLAB), the GlueX \cite{Dobbs:2017vjw} and CLAS12
\cite{Rizzo:2016qvl} experiments are actively searching for more states. At
the ongoing BESIII experiment
\cite{Ablikim:2005um,Kochelev:2005vd,Ablikim:2010au} hybrids can be determined
through decays of charmonia. In the future one expects new insights by the
Panda experiment at FAIR \cite{Lutz:2009ff}.

In the context of flavour multiplets, besides the hybrid meson $\pi_{1}$, one
expects a full nonet of such states. Hence also the kaonic state $K_{1}$ and
two isoscalar states $\eta_{1}$ should exist. In frameworks based on chiral
symmetry an additional nonet of chiral partners should also emerge: these are
so-called pseudovector crypto-exotic hybrid states with quantum numbers
$J^{PC}=1^{+-}$ . Based on the success of chiral models in the ordinary meson
sector, it seems natural to study hybrids in such a framework, in particular
since to our knowledge this has not yet been done before.

In this work, we use the so-called extended Linear Sigma Model (eLSM)
\cite{Parganlija:2012fy,Janowski:2014ppa,Parganlija:2010fz} for this purpose.
Within the eLSM masses and decays of a range of hadrons up to and above $2$
GeV have been described in Refs.~\cite{Parganlija:2012fy,Parganlija:2016yxq}%
.\textbf{ }In particular, in Ref. \cite{Parganlija:2012fy} a fit to various
experimental quantities has shown that a good description of PDG data is
achieved. In addition, this fit allowed to fix univocally the parameters of
the model, and thus to make other predictions/postdictions. Besides
conventional $\bar{q}q$-states, various non-conventional gluonic mesons were
already studied in the eLSM. The scalar glueball appears naturally in the eLSM
as a consequence of dilatation invariance as well as its anomalous breaking.
The resulting dilaton/glueball field mixes with conventional light mesons, and
as shown in Ref.~\cite{Janowski:2014ppa}, is predominantly contained in the
resonance $f_{0}(1710)$. The eLSM has been also applied to the study of the
pseudoscalar glueball(s) \cite{Eshraim:2012jv,Eshraim:2016mds,Eshraim:2019sgr}
and the vector glueball \cite{Giacosa:2016hrm}. Moreover, in the low-energy
domain the eLSM has been shown to be compatible with chiral perturbation
theory \cite{Divotgey:2016pst} for what concerns low-energy pions (most
notably, pion-pion scattering).\ On the other edge, the inclusion of charmed
mesons was presented in\ Refs. \cite{Eshraim:2014eka,Eshraim:2018iea}.

The eLSM has been also successfully applied in the baryonic sector within the
so-called mirror assignment
\cite{Gallas:2009qp,Olbrich:2015gln,Olbrich:2017fsd,Lakaschus:2018rki}, where
pion-nucleon scattering and baryonic decays turn out to be in agreement with
data. One additional advantage of the eLSM is that it can be easily employed
at finite temperature \cite{Kovacs:2016juc,Tawfik:2014gga} and density
\cite{Gallas:2011qp,Heinz:2013hza,Lakaschus:2018rki}, allowing for the
description of the chiral phase transition in the medium.

As discussed in Refs. \cite{Parganlija:2012fy,Janowski:2014ppa}, the general
strategy regarding the Lagrangian construction in eLSM involves implementing
symmetries of relevance for dynamics of low-energy QCD, in particular
dilatation and chiral and dilatation ones and their breaking patters
\cite{Parganlija:2012fy,Janowski:2014ppa}.\textbf{ }Dilatation (or scale)
invariance is a symmetry of the classical QCD Lagrangian that holds in the
chiral limit (that is, when setting the bare masses of the quarks $u,d,s$ to
zero). This symmetry is broken by quantum fluctuations and by the running
coupling which decreases for increasing energy\ (asymptotic freedom), see e.g.
Ref. \cite{Thomas:2001kw}: as a consequence, an energy scale $\Lambda_{QCD}$
emerges. At the hadronic level, these features are described by a
dilaton/glueball field, which mimics the breaking of dilatation invariance
through an appropriate logarithmic potential that involves an energy scale
$\Lambda$ \cite{Migdal:1982jp,Gomm:1985ut}. We assume that -in the chiral
limit and neglecting the second anomaly of QCD, the chiral -or axial- anomaly
(see below)- this is the only way dilatation invariance is broken.\ As a
consequence, in this limit all other interaction terms are dilatation
invariant: this requirement strongly constrains the possible terms that can be
included in the eLSM Lagrangian. Next, it is also required that the eLSM
embodies another key feature of QCD in the low-energy domain: chiral symmetry,
based on the right- and left-handed groups \ $U_{R}(3)\times U_{L}(3),$ and
its spontaneously breaking into $U_{V}(3)$. The pions and kaons appear as
quasi-Goldstone bosons and condensate of scalar fields (scalar $\bar{q}q$
configuration) form in the vacuum. The masses of the chiral partners (such as
scalar and pseudoscalar mesons, but also vector and axial-vector meson, as
well as pseudovector and orbitally excited vector mesons) are not degenerate:
the mass differences are proportional to the afore mentioned chiral
condensates. Finally, terms that are linked to the chiral (or axial anomaly
\cite{tHooft:1986ooh}: this is the second anomaly of QCD, responsible e.g. for
the large mass of the $\eta^{\prime}$ meson) appear: they can break explicitly
the $U_{A=R-L}(1)$ symmetry and/or include the Levi-Civita tensor in the
interaction (Wess-Zumino-type terms \cite{Gomm:1984at}) shall also be added:
they typically also break scale invariance and are important in some decay channels.

In this article, the eLSM setup is extended to hybrids by following the same
strategy related to symmetries outlined above. We construct the chiral
multiplet for the hybrid nonets with $J^{PC}=1^{-+}$ and $J^{PC}=1^{+-}$ and
determine the interaction terms which satisfy chiral symmetry. As a
consequence, the spontaneous breaking of chiral symmetry is responsible for
the mass differences between the low-lying $1^{-+}$ exotic hybrids and the
heavier $1^{+-}$ crypto-exotic hybrids, just as among standard quark-antiquark
chiral partners. The possible decays of the hybrids in the two multiplets are
described by four interaction terms. Two of these fulfill chiral dilatation
invariance and therefore should be dominant. The third and the fourth terms
break dilatation invariance and are linked to the chiral anomaly. In
particular, the third term involves the Levi-Civita tensor and the fourth term
breaks explicitly the axial anomaly $U(1)_{A}$. We work out the resulting
decays and identify promising channels for the experimental discovery of these states.

As mentioned above, two hybrid candidates $\pi_{1}(1400)$ and $\pi_{1}(1600)$
are listed in the PDG \cite{Tanabashi:2018oca}. However, in the recent
theoretical analysis of Ref.~\cite{Rodas:2018owy} it was suggested that these
two states could correspond to a single resonant pole, with mass and width
close to the original $\pi_{1}(1600)$. Indeed, our chiral multiplet -just as
other models and lattice studies- has space for only one such $\pi_{1}$-state:
we then adopt the interpretation of Ref.~\cite{Rodas:2018owy} and use the mass
of the $\pi_{1}(1600)$ as an input that fixes the masses of hybrids in our framework.

This paper is organized as follows. In Sec.~\ref{sec:chiral} we present the
standard quark-antiquark nonets in the eLSM and construct the new hybrid
nonets and their transformation properties. In Sec.~\ref{sec:Lagrangian} we
introduce the effective Lagrangian and discuss the interaction terms that lead
to the hybrid decays. In Sec.~\ref{sec:results} we present and discuss our
results and in Sec.~\ref{sec:conclusions} we outline our conclusions and
outlook. Technical details of our calculations are relegated to several appendices.

\section{Chiral multiplets}

\label{sec:chiral}

In this section, we briefly review the assignment of (pseudo)scalar,
(axial-)vector and pseudovector fields, which are the basic ingredients of the
eLSM. Then, we show how to build two nonets of hybrid states with quantum
numbers $J^{PC}=1^{+-}$ and $J^{PC}=1^{-+}$.

\subsection{(Pseudo)scalar and (axial-)vector quark-antiquark multiplets}

The nonets of (pseudo)scalar fields are introduced as
\begin{equation}
P=\frac{1}{\sqrt{2}}\left(
\begin{array}
[c]{ccc}%
\frac{\eta_{N}+\pi^{0}}{\sqrt{2}} & \pi^{+} & K^{+}\\
\pi^{-} & \frac{\eta_{N}-\pi^{0}}{\sqrt{2}} & K^{0}\\
K^{-} & \bar{K}^{0} & \eta_{S}%
\end{array}
\right)  \text{ , }S=\frac{1}{\sqrt{2}}\left(
\begin{array}
[c]{ccc}%
\frac{\sigma_{N}+a_{0}^{0}}{\sqrt{2}} & a_{0}^{+} & K_{S}^{+}\\
a_{0}^{-} & \frac{\sigma_{N}-a_{0}^{0}}{\sqrt{2}} & K_{S}^{0}\\
K_{S}^{-} & \bar{K}_{S}^{0} & \sigma_{S}%
\end{array}
\right)  \text{ .}%
\end{equation}
The matrix $P$ contains the light pseudoscalar nonet $\{\pi$, $K,\eta
,\eta^{\prime}\}$ with quantum numbers $J^{PC}=0^{-+}$
\cite{Tanabashi:2018oca}, where $\eta$ and $\eta^{\prime}$ arise via the
mixing $\eta=\eta_{N}\cos\theta_{p}+\eta_{S}\sin\theta_{p},$ $\eta^{\prime
}=-\eta_{N}\sin\theta_{p}+\eta_{S}\cos\theta_{p}$ with $\theta_{p}%
\simeq-44.6^{\circ}$ \cite{Parganlija:2012fy}. Using other values for the
mixing angle such as $\theta_{p}=-41.4^{\circ}$ \cite{AmelinoCamelia:2010me}
changes only slightly the results presented in this work. The matrix $S$
contains the scalar fields \{$a_{0}(1450),$ $K_{0}^{\ast}(1430),$ $\sigma
_{N},$ $\sigma_{S}$\} with $J^{PC}=0^{++}$. These are identified with states
above $1$ GeV \cite{Parganlija:2012fy}: the non-strange bare field $\sigma
_{N}\equiv\left\vert \bar{u}u+\bar{d}d\right\rangle /\sqrt{2}$ corresponds
predominantly to the resonance \thinspace$f_{0}(1370)$ and the bare field
$\sigma_{S}\equiv\left\vert \bar{s}s\right\rangle $ predominantly to
$f_{0}(1500)$. As already indicated above, the state $f_{0}(1710)$ is
dominated by the scalar glueball. For details of the mixing see
Ref.~\cite{Janowski:2014ppa}. Evidence for a large gluonic component in
$f_{0}(1710)$ has also been found on the lattice \cite{Gui:2012gx}
and in the holographic QCD study of
Refs.~\cite{Brunner:2015oqa,Brunner:2015yha,Brunner:2015oga}.

In the eLSM, the nonet of the light scalar states \{$a_{0}(980),$ $K_{0}%
^{\ast}(700),$ $f_{0}(500),$ $f_{0}(980)$\} turns out to be non-$q\bar{q}$.
One possibility is a nonet of light tetraquark states
\cite{Jaffe:2004ph,Pelaez:2003dy,Napsuciale:2004au,Fariborz:2003uj,Fariborz:2005gm,Maiani:2004uc,Giacosa:2006rg,
Giacosa:2009qh,Heupel:2012ua,Eichmann:2015cra,Pelaez:2015qba} and/or a nonet
of dynamically generated states
\cite{vanBeveren:1986ea,Oller:1997ti,Oller:1998hw,vanBeveren:2006ua}).
Moreover, these two configurations can mix with each other, making a clear
distinction quite difficult. Nevertheless, there is an agreement toward the
interpretation of the light scalar nonet as a nonet of states made up with
four quarks.

The scalar and pseudoscalar matrices are combined into the matrix%
\begin{equation}
\Phi=S+iP\text{ ,} \label{phimat}%
\end{equation}
which has a simple transformation under chiral transformations $U_{L}(3)\times
U_{R}(3)$: $\Phi\rightarrow U_{L}\Phi U_{R}^{\dagger}$, where $U_{L}$ and
$U_{R}$ are unitary $U(3)$ matrices. Under parity $P$ the matrix $\Phi$
transforms as $\Phi\rightarrow\Phi^{\dagger}$ and under charge conjugation $C$
as $\Phi\rightarrow\Phi^{t}$. The matrix $\Phi$ is used as a building block in
the construction of the eLSM Lagrangian, see Appendix \ref{appA} and Tables
\ref{tabI} and \ref{tabII}. For a detailed report of the transformation
properties, we refer to Ref. \cite{Parganlija:2012xj}.

A comment on the currents is in order: the quantity $\bar{q}i\gamma^{5}q$ is a
pseudoscalar ($J^{PC}=0^{-+}$), as a simple check on the transformations at
the quark level shows. Then, in the framework of a nonrelativistic approach it
corresponds to $L=0$ and $S=0,$ for which $P=(-1)^{L+1}=-1$ and $C=(-1)^{L+S}%
=+1.$ One may also (and in more detail) verify this correspondence by studying
$\bar{q}i\gamma^{5}q$ in the nonrelativistic limit, by taking the dominant
components within the Dirac representation for the spinors. For what concerns
the object $\bar{q}q,$ this is clearly a scalar object ($J^{PC}=0^{++}$); in
the non-relativistic language, it is obtained by choosing $L=1$ and $S=1$
coupled to $J=0.$ Also in this case, even if not so obvious at a first sight,
a nonrelativistic decomposition of $\bar{q}q$ shows that $L=1$ and the spin
triplet configuration emerge.

We now turn to vector and axial-vector fields, described by:
\begin{equation}
V^{\mu}=\frac{1}{\sqrt{2}}\left(
\begin{array}
[c]{ccc}%
\frac{\omega_{N}+\rho^{0}}{\sqrt{2}} & \rho^{+} & K^{\star+}\\
\rho^{-} & \frac{\omega_{N}-\rho^{0}}{\sqrt{2}} & K^{\star0}\\
K^{\star-} & \bar{K}^{\star0} & \omega_{S}%
\end{array}
\right)  ^{\mu}\;\text{, }A^{\mu}=\frac{1}{\sqrt{2}}\left(
\begin{array}
[c]{ccc}%
\frac{f_{1N}+a_{1}^{0}}{\sqrt{2}} & a_{1}^{+} & K_{1,A}^{+}\\
a_{1}^{-} & \frac{f_{1N}-a_{1}^{0}}{\sqrt{2}} & K_{1,A}^{0}\\
K_{1,A}^{-} & \bar{K}_{1,A}^{0} & f_{1S}%
\end{array}
\right)  ^{\mu}\;\text{.} \label{noneta}%
\end{equation}
The elements of the matrix $V^{\mu}$ are the vector states \{$\rho(770),$
$K^{\ast}(892),$ $\omega(782),$ $\phi(1020)$\} with $J^{PC}=1^{+-}$, and the
elements of the matrix $A^{\mu}$ the axial-vector states \{$a_{1}(1230),$
$K_{1,A},$ $f_{1}(1285),$ $f_{1}(1420)$\} with $J^{PC}=1^{++}$. Here,
$K_{1,A}$ is a mixture of the two physical states $K_{1}(1270)$ and
$K_{1}(1400),$ see also Sec.~\ref{sec:pseudovector}. We neglect (the anyhow
small) strange-nonstrange mixing, hence $\omega_{N}\equiv\omega(782)$ and
$f_{1N}\equiv f_{1}(1285)$ are regarded as purely nonstrange mesons of the
type $\sqrt{1/2}(\bar{u}u+\bar{d}d)$, while $\omega_{S}\equiv\phi(1020)$ and
$f_{1S}\equiv f_{1}(1420)$ are regarded as purely $\bar{s}s$ states.

Next, one defines the right-handed and left-handed combinations:
\begin{equation}
R^{\mu}=V^{\mu}-A^{\mu}\text{ and }L^{\mu}=V^{\mu}+A^{\mu}\text{ .} \label{rl}%
\end{equation}
Under chiral transformation they transform as $R^{\mu}\rightarrow U_{R}R^{\mu
}U_{R}^{\dagger}$ and $L^{\mu}\rightarrow U_{L}L^{\mu}U_{L}^{\dagger}$.
Details of the currents and transformations are shown in Tables \ref{tabI} and
\ref{tabII}.

The vector current $\bar{q}\gamma^{\mu}q$ with $J^{PC}=1^{--}$ corresponds to
$L=1$ and $S=0,$ while the axial-vector current $\bar{q}\gamma^{5}\gamma^{\mu
}q$ with $J^{PC}=1^{++}$to $L=1$ and $S=1,$ coupled to $J=1.$ Again, these
correspondences can be also verified by studying the nonrelativistic limits of
the quark-antiquark currents $\bar{q}\gamma^{\mu}q$ and $\bar{q}\gamma
^{5}\gamma^{\mu}q$.

The eLSM Lagrangian includes the multiplets $S,$ $P,$ $V,$ and $A$ presented
above. In addition, a dilaton/glueball field is also present in order to
describe dilatation symmetry and its anomalous breaking. The details of the
eLSM (together with its symmetries, most notably chiral and dilatation
symmetries and their anomalous, explicit, and spontaneous breaking terms) are
briefly summarized in Appendix \ref{appA} and extensively presented in Refs.
\cite{Parganlija:2012fy,Janowski:2014ppa} for $N_{f}=3$. An extension to
$N_{f}=4$ can be found in Refs.~\cite{Eshraim:2014eka,Eshraim:2018iea} and a
study of mesons at finite temperature can be found in
Refs.~\cite{Kovacs:2016juc,Tawfik:2014gga}.

\subsection{Pseudo-vector and excited vector mesons}

\label{sec:pseudovector}

Since we are interested in hybrids, it is important to consider also the
pseudo-vector mesons with quantum numbers $J^{PC}=1^{+-}$ and the excited
vector mesons with quantum numbers $J^{PC}=1^{--}$ , since they are important
decay products of hybrids. To this end we introduce the matrices (see Ref.
\cite{Giacosa:2016hrm} for technical details):
\begin{equation}
B^{\mu}=\frac{1}{\sqrt{2}}%
\begin{pmatrix}
\frac{h_{1,N}+b_{1}^{0}}{\sqrt{2}} & b_{1}^{+} & K_{1,B}^{\star+}\\
b_{1}^{-} & \frac{h_{1,N}+b_{1}^{0}}{\sqrt{2}} & K_{1,B}^{\star0}\\
K_{1,B}^{\star-} & \bar{K}_{1,B}^{\star0} & h_{1,S}%
\end{pmatrix}
^{\mu}\text{ , }V_{E}^{\mu}=\frac{1}{\sqrt{2}}%
\begin{pmatrix}
\frac{\omega_{E,N}+\rho_{E}^{0}}{\sqrt{2}} & \rho_{E}^{+} & K_{E}^{\star+}\\
\rho_{E}^{-} & \frac{\omega_{E,N}-\rho_{E}^{0}}{\sqrt{2}} & K_{E}^{\star0}\\
K_{E}^{\star-} & \bar{K}_{E}^{\star0} & \omega_{E,S}%
\end{pmatrix}
^{\mu}\,. \label{nonetb}%
\end{equation}
Here, $B^{\mu}$ contains the pseudovector states $\{b_{1}(1230),K_{1,B}%
,h_{1}(1170),h_{1}(1380)\}$. In the quark model these states emerge from
$L=1,$ $S=0$ coupled to $J^{PC}=1^{+-}$ (hence, pseudovector states as
axial-vector states with negative $C$-parity). For simplicity, the
strange-nonstrange isoscalar mixing is again neglected, thus $h_{1,N}\equiv
h_{1}(1170)$ is a purely nonstrange state, while $h_{1,S}\equiv h_{1}(1380)$
is a purely strange-antistrange state. Note, these states are distinguished
from the axial-vector states of Eq.~(\ref{noneta}) due to $C$-parity. However,
$C$-parity does not apply for kaonic states and mixing arises. The kaonic
fields $K_{1,A}$ from Eq. (\ref{noneta}) and $K_{1,B}$ from Eq. (\ref{nonetb})
mix and generate the two physical resonances $K_{1}(1270)$ and $K_{1}(1400)$:%
\begin{equation}%
\begin{pmatrix}
K_{1}^{+}(1270)\\
K_{1}^{+}(1400)
\end{pmatrix}
^{\mu}=%
\begin{pmatrix}
\cos\varphi & -i\sin\varphi\\
-i\sin\varphi & \cos\varphi
\end{pmatrix}%
\begin{pmatrix}
K_{1,A}^{+}\\
K_{1,B}^{+}%
\end{pmatrix}
^{\mu}\text{ .} \label{mixk}%
\end{equation}
The mixing angle reads $\varphi=(56.3\pm4.2)^{\circ}$ \cite{Divotgey:2013jba}.
The same transformations hold for $K_{1}^{0}(1270)$ and $K_{1}^{0}(1400),$
while for the other kaonic states one has to take into account that $K_{1}%
^{-}(1270)=K_{1}^{+}(1270)^{\dagger}$ and $\bar{K}_{1}^{0}(1270)=K_{1}%
^{0}(1270)^{\dagger}$ (and so for $K_{1}^{-}(1400)$). Notice that the
imaginary number $i$ is a consequence of the specific mixing term between the
fields $K_{1,A}^{+}$ and $K_{1,B}^{+}$ fields, that must be invariant under
charge-conjugation and parity, and the specific convention used in Ref.
\cite{Divotgey:2013jba}. According to this setup, under $C$-transformation the
field $K_{1}^{+}(1270)$ changes into $K_{1}^{-}(1270),$ while $K_{1}%
^{+}(1400)$ into $-K_{1}^{-}(1400)$.\ Since the kaonic fields are not
eigenstates of the charge conjugation operator $C$, other choices are
possible, which however do not change the physical results.

The chiral partners of the pseudovector mesons are excited vector mesons which
arise from the combination $L=2,$ $S=1$ coupled to $J^{PC}=1^{--}$. The
corresponding fields listed are given by $\{\rho(1700)\text{, }K^{\ast
}(1680),\text{ }\omega(1650),\text{ }\phi(1930?)\}$. The experimental evidence
of the first three states is compiled by the PDG, while the putative new state
$\phi(1930?)$ is expected to couple predominantly to $K$ and $K^{\ast}$
according to the study of Ref.~\cite{Piotrowska:2017rgt}. The question mark in
$\phi(1930?)$ means that presently this state (and the corresponding mass of
$1930$ MeV) is only a theoretical prediction.

We then build the matrix
\begin{equation}
\tilde{\Phi}^{\mu}=V_{E}^{\mu}-iB^{\mu}\text{ ,}%
\end{equation}
which under chiral transformations changes as $\tilde{\Phi}^{\mu}\rightarrow
U_{L}\tilde{\Phi}^{\mu}U_{R}^{\dag}$ (it is a so-called heterochiral
multiplet, just as the standard (pseudo)scalar $\Phi$), under parity as
$\tilde{\Phi}^{\mu}\rightarrow\tilde{\Phi}^{\dag\mu}$, and under charge
conjugations as $\tilde{\Phi}^{\mu}\rightarrow-\tilde{\Phi}^{t,\mu},$ see
Tables \ref{tabI} and \ref{tabII} for details. As shown in
Ref.~\cite{Giacosa:2017pos}, further chiral multiplets can be built in an
analogous way.

The currents for these fields involve derivatives. The pseudovector current
$\bar{q}i\gamma^{5}\overleftrightarrow{\partial}^{\mu}q$ with $J^{PC}=1^{+-}$
corresponds to $L=1$ and $S=0.$ Intuitively, it is obtained from $\bar
{q}i\gamma^{5}q$ (with $L=0,$ $S=0)$ by adding a derivative, which increases
the angular momentum of one unit, hence $L=1.$ Similarly, the excited vector
current $\bar{q}\overleftrightarrow{\partial}^{\mu}q$ with $J^{PC}=1^{--}$
corresponds to $L=2$ and $S=1$ coupled to $J=1.$ Again, it is obtained from
the object $\bar{q}q$ (which has $L=S=1)$ upon adding the derivative, lifting
$L$ to $2.$ Thus, this is the nonet of orbitally excited vector mesons. Such
statements can be checked, just as in the previous cases, by a nonrelativistic
study of the quark-antiquark currents $\bar{q}i\gamma^{5}\overleftrightarrow
{\partial}^{\mu}q$ and $\bar{q}\overleftrightarrow{\partial}^{\mu}q$.

\begin{table}[h]
\centering
\begin{tabular}
[c]{|c|c|c|c|c|c|}\hline
Nonet & $J^{PC}$ & Current & Assignment\rule{0pt}{3.7ex}\rule[-2ex]{0pt}{0pt}
& $P$ & $C$\\\hline
$P$ & $0^{-+}$ & $P_{ij}=\frac{1}{\sqrt{2}}\bar{q}_{j}i\gamma^{5}q_{i}$ &
\rule{0pt}{3.7ex}\rule[-2ex]{0pt}{0pt} $\pi,K,\eta,\eta^{\prime}$ &
$-P(t,-\mathbf{x})$ & $P^{t}$\\\hline
$S$ & $0^{++}$ & $S_{ij}=\frac{1}{\sqrt{2}}\bar{q}_{j}q_{i}$ & $a_{0}(1450),$
$K_{0}^{\ast}(1430),$ $f_{0}(1370),$ $f_{0}(1510)$ \rule{0pt}{3.7ex}%
\rule[-2ex]{0pt}{0pt} & $S(t,-\mathbf{x})$ & $S^{t}$\\\hline
$V^{\mu}$ & $1^{--}$ & $V_{ij}^{\mu}=\frac{1}{\sqrt{2}}\bar{q}_{j}\gamma^{\mu
}q_{i}$ & $\rho(770),$ $K^{\ast}(892),$ $\omega(785)$, $\phi(1024)$
\rule{0pt}{3.7ex}\rule[-2ex]{0pt}{0pt} & $V_{\mu}(t,-\mathbf{x})$ &
$-V^{\mu,t}$\\\hline
$A^{\mu}$ & $1^{++}$ & $A_{ij}^{\mu}=\frac{1}{\sqrt{2}}\bar{q}_{j}\gamma
^{5}\gamma^{\mu}q_{i}$ & $a_{1}(1230),$ $K_{1,A},$ $f_{1}(1285),$
$f_{1}(1420)$ \rule{0pt}{3.7ex}\rule[-2ex]{0pt}{0pt} & $-A_{\mu}%
(t,-\mathbf{x})$ & $A^{\mu,t}$\\\hline
$B^{\mu}$ & $1^{+-}$ & $B_{ij}^{\mu}=\frac{1}{\sqrt{2}}\bar{q}_{j}\gamma
^{5}\overleftrightarrow{\partial}^{\mu}q_{i}$ & $b_{1}(1230),\text{ }%
K_{1,B},\text{ }h_{1}(1170),\text{ }h_{1}(1380)$ & $-B_{\mu}(t,-\mathbf{x})$ &
$-B^{\mu,t}$ \rule{0pt}{3.7ex}\rule[-2ex]{0pt}{0pt}\\\hline
$V_{E}^{\mu}$ & $1^{--}$ & $V_{E,ij}^{\mu}=\frac{1}{\sqrt{2}}\bar{q}%
_{j}i\overleftrightarrow{\partial}^{\mu}q_{i}$ & $\rho(1700),$ $K^{\ast
}(1680),$ $\omega(1650)$, $\phi(1930?)$ \rule{0pt}{3.7ex}\rule[-2ex]{0pt}{0pt}
& $V_{\mu}(t,-\mathbf{x})$ & $-V_{E}^{\mu,t}$\\\hline
$\Pi^{hyb,\mu}$ & $1^{-+}$ & $\Pi_{ij}^{hyb,\mu}=\frac{1}{\sqrt{2}}\bar{q}%
_{j}G^{\mu\nu}\gamma_{\nu}q_{i}$ & $\pi_{1}(1600),K_{1}(?),\eta_{1}%
(?),\eta_{1}(?)$ & $\Pi_{\mu}^{hyb}(t,-\mathbf{x})$ & $\Pi^{hyb,\mu,t}%
$\rule{0pt}{3.7ex}\rule[-2ex]{0pt}{0pt}\\\hline
$B^{hyb,\mu}$ & $1^{+-}$ & $B_{ij}^{hyb,\mu}=\frac{1}{\sqrt{2}}\bar{q}%
_{j}G^{\mu\nu}\gamma_{\nu}\gamma^{5}q_{i}$ & $b_{1}(2000?),$ $K_{1,B}(?),$
$h_{1}(?),$ $h_{1}(?)$ & $-B_{\mu}^{hyb}(t,-\mathbf{x})$ & $-B^{hyb,\mu,t}%
$\rule{0pt}{3.7ex}\rule[-2ex]{0pt}{0pt}\\\hline
\end{tabular}
\caption{Summary of the quark-antiquark and hybrid nonets and their
properties.}%
\label{tabI}%
\end{table}

\begin{table}[h]
\centering
\begin{tabular}
[c]{|c|c|c|c|c|}\hline
Chiral multiplet & Current & $U_{R}(3)\times U_{L}(3)$ & $P$ & $C$%
\rule{0pt}{3.7ex}\rule[-2ex]{0pt}{0pt}\\\hline
$\Phi=S+iP$ & $\sqrt{2}\bar{q}_{R,j}q_{L,i}$ & $U_{L}\Phi U_{R}^{\dagger}$ &
$\Phi^{\dag}$ & $\Phi^{t}$ \rule{0pt}{3.7ex}\rule[-2ex]{0pt}{0pt}\\\hline
$R^{\mu}=V^{\mu}-A^{\mu}$ & $\sqrt{2}\bar{q}_{R,j}\gamma^{\mu}q_{R,i}$ &
$U_{R}R^{\mu}U_{R}^{\dagger}$ & $L_{\mu}$ & $-\left(  L^{\mu}\right)  ^{t}%
$\rule{0pt}{3.7ex}\rule[-2ex]{0pt}{0pt}\\\hline
$L^{\mu}=V^{\mu}+A^{\mu}$ & $\sqrt{2}\bar{q}_{L,j}\gamma^{\mu}q_{L,i}$ &
$U_{L}R^{\mu}U_{L}^{\dagger}$ & $R_{\mu}$ & $-\left(  R^{\mu}\right)  ^{t}%
$\rule{0pt}{3.7ex}\rule[-2ex]{0pt}{0pt}\\\hline
$\tilde{\Phi}^{\mu}=V_{E}^{\mu}-iB^{\mu}$ & $\sqrt{2}\bar{q}_{R,j}%
i\overleftrightarrow{\partial}^{\mu}q_{L,i}$ & $U_{L}\tilde{\Phi}^{\mu}%
U_{R}^{\dagger}$ & $\tilde{\Phi}_{\mu}^{\dag}$ & $-\tilde{\Phi}^{t\mu}%
$\rule{0pt}{3.7ex}\rule[-2ex]{0pt}{0pt}\\\hline
$R^{hyb,\mu}=\Pi^{hyb,\mu}-B^{hyb,\mu}$ & $\sqrt{2}\bar{q}_{R,j}G^{\mu\nu
}\gamma_{\nu}q_{R,i}$ & $U_{R}R^{hyb,\mu}U_{R}^{\dagger}$ & $L_{\mu}^{hyb}$ &
$(L^{hyb,\mu}\rule{0pt}{3.7ex}\rule[-2ex]{0pt}{0pt})^{t}$\rule{0pt}{3.7ex}%
\rule[-2ex]{0pt}{0pt}\\\hline
$L^{hyb,\mu}=\Pi^{hyb,\mu}+B^{hyb,\mu}$ & $\sqrt{2}\bar{q}_{L,j}G^{\mu\nu
}\gamma_{\nu}q_{L,i}$ & $U_{L}R^{hyb,\mu}U_{L}^{\dagger}$ & $R_{\mu}^{hyb}$ &
$(R^{hyb,\mu}\rule{0pt}{3.7ex}\rule[-2ex]{0pt}{0pt})^{t}$\rule{0pt}{3.7ex}%
\rule[-2ex]{0pt}{0pt}\\\hline
\end{tabular}
\caption{Transformation properties of the chiral multiplets.}%
\label{tabII}%
\end{table}

\subsection{Hybrid multiplets}

In this subsection, we introduce hybrids. The currents of exotic hybrid states
with quantum numbers $J^{PC}=1^{-+}$ are given by
\begin{equation}
\Pi_{ij}^{hyb,\mu}=\frac{1}{\sqrt{2}}\bar{q}_{j}G^{\mu\nu}\gamma_{\nu}%
q_{i}\text{ ,} \label{hyb1}%
\end{equation}
where $G^{\mu\nu}=\partial^{\mu}A^{\nu}-\partial^{\mu}A^{\nu}-g_{QCD}[A^{\mu
},A^{\nu}]$ is the gluonic field tensor. Thus, these currents can be
understood as `vector currents with the addition of one gluon', which is
responsible for the switch of the $C$-parity. Note, the emerging quantum
numbers are exotic (not allowed for a local quark-antiquark current).
According to lattice QCD, these are the lightest hybrid states
\cite{Michael:1999ge,Dudek:2009qf,Dudek:2010wm,Dudek:2013yja}.

In other words, it is important to stress there is no way to build a local
current with $J^{PC}=1^{-+}$ by combining a quark and an antiquark, such as
$\bar{q}\Gamma q$ where $\Gamma$ is a combination of Dirac matrices and/or
derivatives (see the previous subsection for specific examples). In the
nonrelativistic language, one cannot construct a $\bar{q}q$ object with
$J^{PC}=1^{-+}$ [the conditions $P=(-1)^{L+1}$ and $C=(-1)^{L+S}$ together
with $J=\left\vert L-S\right\vert ,...,L+S$ cannot be fulfilled
simultaneously]. Nevertheless, besides having exotic quantum numbers, we still
have a nonet of states, just as for a regular quark-antiquark nonet. Thus, for
what concerns the construction of effective models, the nonet $\Pi^{hyb,\mu}$
can be used as a building block of the model.

The chiral partners of $\Pi_{ij}^{hyb,\mu}$ are the pseudo-vector states
$B_{ij}^{hyb,\mu}$, which have the quantum numbers $J^{PC}=1^{+-}$ and are
given by%
\begin{equation}
B_{ij}^{hyb,\mu}=\frac{1}{\sqrt{2}}\bar{q}_{j}G^{\mu\nu}\gamma^{5}\gamma_{\nu
}q_{i}\text{ .} \label{hyb2}%
\end{equation}
The quantum numbers are the same as for pseudovector mesons, even if the
underlying currents are utterly different. We thus refer to these states as
crypto-exotic mesons, since we do have a nonet of non-conventional hybrid
states, but the quantum numbers are also allowed for normal quark-antiquark
states (see above). A mixing of configurations with the same quantum numbers
is in principle possible. Yet, in the present case the mass differences
between conventional $\bar{q}q$ pseudovector and hybrid mesons are large
enough to neglect such a mixing.

In terms of matrices, we have%
\begin{equation}
\Pi^{hyb,\mu}=\frac{1}{\sqrt{2}}\left(
\begin{array}
[c]{ccc}%
\frac{\eta_{1,N}^{hyb}+\pi_{1}^{0}}{\sqrt{2}} & \pi_{1}^{hyb+} & K_{1}%
^{hyb+}\\
\pi_{1}^{hyb-} & \frac{\eta_{1,N}^{hyb}-\pi_{1}^{0}}{\sqrt{2}} & K_{1}%
^{hyb0}\\
K_{1}^{hyb-} & \bar{K}_{1}^{hyb0} & \eta_{1,S}^{hyb}%
\end{array}
\right)  ^{\mu}\;\text{, }B^{hyb,\mu}=\frac{1}{\sqrt{2}}\left(
\begin{array}
[c]{ccc}%
\frac{h_{1N,B}^{hyb}+b_{1}^{hyb,0}}{\sqrt{2}} & b_{1}^{hyb,+} & K_{1,B}%
^{hyb+}\\
b_{1}^{hyb,+} & \frac{h_{1N,B}^{hyb}-b_{1}^{hyb,0}}{\sqrt{2}} & K_{1,B}%
^{hyb0}\\
K_{1,B}^{hyb-} & \bar{K}_{1,B}^{hyb0} & h_{1S,B}^{hyb}%
\end{array}
\right)  ^{\mu}\text{ .} \label{hybridmatrices}%
\end{equation}
For the hybrid states contained in $\Pi^{hyb,\mu}$, the field $\pi_{1}^{hyb}$
is assigned to $\pi_{1}(1600)$ \cite{Rodas:2018owy}, as already discussed in
the introduction. For the other members of the nonet, no experimental
candidates are yet known. In Sec.~\ref{sec:results} we will present our
estimate for their masses and decays.

For the chiral partners contained in $B^{hyb,\mu}$ again no candidate exists.
In a lattice simulation no states below $2.4$ GeV have been found
\cite{Dudek:2010wm}, but this result has to be interpreted with caution due to
the large pion masses (about $400$ MeV) used in the simulation. We estimate
the mass of the chiral partner of $\pi_{1},$ the so-called $b_{1}^{hyb}$
state, to have a mass in the $2$-$2.5$ GeV range, once that the pion mass
converges to the physical value. For definiteness, we shall assign it to an
hypothetical state to the lower limit $b_{1}(2000?)$ state, but our results do
not change much when increasing this mass up to $2.4$ GeV. The masses of the
other members of the pseudovector crypto-exotic nonet then follow as a
consequence of this assumption. [In this work we restrict to hybrids
containing only the light quarks $u,d,$ and $s.$ It should be also stressed
that hybrid mesons can be also realized with heavy quarks, see the lattice
studies in\ Refs. \cite{Liu:2012ze,Moir:2013ub}: the corresponding currents
have a similar form as in\ Eq. (\ref{hyb1}) and (\ref{hyb2}).]

For completeness, in the Tables \ref{tabI} and \ref{tabII} we summarize all
relevant properties and transformations of the nonets introduced in this section.

\section{The Lagrangian terms involving hybrid mesons}

\label{sec:Lagrangian}

In this section we present the enlarged eLSM Lagrangian involving hybrids. We
start form the general form
\begin{equation}
\mathcal{L}_{eLSM}^{\text{enlarged}}=\mathcal{L}_{eLSM}+\mathcal{L}%
_{eLSM}^{\text{ hybrid }}%
\end{equation}
where $\mathcal{L}_{eLSM}$ is the standard part, built under chiral and
dilatation symmetries, as well as their spontaneous and explicit breaking
features (see Appendix \ref{appA}). Next, the hybrid part is written as:%

\begin{equation}
\mathcal{L}_{eLSM}^{\text{ hybrid }}=\mathcal{L}_{eLSM}^{\text{
hybrid-quadratic}}+\mathcal{L}_{eLSM}^{\text{ hybrid-linear}}\text{ .}%
\end{equation}
These terms will be discussed separately in this section. Before doing so, a
general comment is in order: In this work, the masses and decays are
calculated at tree-level.\ This is in agreement with the basic strategy of the
eLSM, which amounts to include as much as possible mesonic (and baryonic)
interpolating fields. As a matter of fact, this strategy has shown to be quite
successful in describing some aspects of the low-energy QCD phenomenology
\cite{Parganlija:2012fy,Olbrich:2015gln}. Namely, we recall that in the
mesonic sector it was possible to describe the meson phenomenology up to $1.7$
GeV. In the extension of Ref. \cite{Parganlija:2016yxq} also mesons above 2
GeV have been considered. We are therefore confident that a linear chiral
model, which by construction includes chiral partners, can give useful results
also for the yet unknown hybrid states (one nonet at about 1.7 GeV and one
just above 2 GeV), that are the subject of the present work.

Of course, as a matter of principle, the necessity of unitarization and its
influence on statements made in this paper represent a relevant question. To
this end, we note that in Ref. \cite{Giacosa:2007bn} it was shown that, as
long as the width-to-mass ratio $\Gamma/M$ is sufficiently small (smaller than
$0.2$, the ratio in the case of the $\rho$ meson) the effect of loops -which
is large-$N_{c}$ suppressed- is not expected to change the picture; later on,
in\ Ref. \cite{Schneitzer:2014rsa} the next-to-leading triangle diagram for
two-body decays has been shown to be negligible.

On the other hand, there are indeed some cases where the role of loops can be
important.\ This is generally true for scalar mesons. For instance, in the
eLSM the $a_{0}(1450)$ is a predominantly a $\bar{q}q$ state, and the
$a_{0}(980)$ is not part of the model. When including loops it is possible to
show that the $a_{0}(980)$ emerges as an additional companion pole as a kind
of four-quark state \cite{Wolkanowski:2015lsa} (see also Ref.
\cite{Boglione:2002vv}). Similarly, the resonance $K_{0}^{\ast}(700)$ is
dynamically generated through loops and is the companion state of the
predominantly $\bar{q}q$ state $K_{0}^{\ast}(1430)$ \cite{Wolkanowski:2015jtc}%
, see also Ref. \cite{Black:2000qq}.

In conclusion, while loops are relevant for (relatively) broad resonances, the
tree-level results represent a clear and well-definite set-up to get
meaningful results such as decay ratios. The inclusion of loop effects should
be performed in the future when a better experimental knowledge will be available.

\subsection{Quadratic terms in the hybrid fields hybrid kinetic terms and
masses}

The quadratic term for the hybrid fields can be decomposed as
\begin{equation}
\mathcal{L}_{eLSM}^{\text{ hybrid-quadratic}}=\mathcal{L}_{eLSM}^{\text{
hybrid-kin}}+\mathcal{L}_{eLSM}^{\text{ hybrid-mass}}\text{ ,}%
\end{equation}
where one has the usual vectorial kinetic term
\begin{equation}
\mathcal{L}_{eLSM}^{\text{ hybrid-kin}}=-\mathrm{Tr}\left(  L_{\mu\nu}%
^{hyb,2}+R_{\mu\nu}^{hyb,2}\right)  =-\mathrm{Tr}\left(  V_{\mu\nu}%
^{hyb,2}+A_{\mu\nu}^{hyb,2}\right)  \,,
\end{equation}
with
\begin{equation}
V_{\mu\nu}^{hyb}=\partial_{\mu}V_{\nu}^{hyb}-\partial_{\nu}V_{\mu}^{hyb}\text{
and }A_{\mu\nu}^{hyb}=\partial_{\mu}A_{\nu}^{hyb}-\partial_{\nu}A_{\mu}^{hyb}.
\end{equation}
Moreover, we consider the term describing the masses of hybrids as
\begin{align}
\mathcal{L}_{eLSM}^{\text{ hybrid-mass}}=  &  m_{1}^{hyb,2}\frac{G^{2}}%
{G_{0}^{2}}\mathrm{Tr}\left(  L_{\mu}^{hyb,2}+R_{\mu}^{hyb,2}\right)
+\mathrm{Tr}\left(  \Delta^{hyb}\left(  L_{\mu}^{hyb,2}+R_{\mu}^{hyb,2}%
\right)  \right) \nonumber\\
&  +\frac{h_{1}^{hyb}}{2}\mathrm{Tr}(\Phi^{\dagger}\Phi)\mathrm{Tr}\left(
L_{\mu}^{hyb,2}+R_{\mu}^{hyb,2}\right)  +h_{2}^{hyb}\mathrm{Tr}\left[
\left\vert L_{\mu}^{hyb}\Phi\right\vert ^{2}+\left\vert \Phi R_{\mu}%
^{hyb}\right\vert ^{2}\right]  +2h_{3}^{nyb}\mathrm{Tr}\left(  L_{\mu}%
^{hyb}\Phi R^{hyb,\mu}\Phi^{\dagger}\right)  \text{ ,} \label{lelsmhybquad}%
\end{align}
which fulfills both chiral and dilatation invariances. Note, the dilaton field
$G$ as well as its vacuum's expectation value $G_{0}$ enter into these
expressions, see Appendix \ref{appA} and Refs.
\cite{Parganlija:2012fy,Janowski:2014ppa,Parganlija:2010fz}.

The masses of hybrids can be calculated from the previous expressions by
taking into account that the (pseudo)scalar field $\Phi$ has a nonzero
condensate or vacuum's expectation value (v.e.v.): $\Phi\equiv\Phi
_{0}=diag\{\phi_{N}/2,\phi_{N}/2,\phi_{S}/\sqrt{2}\}.\ $This condensate
reflects the spontaneous breaking of chiral symmetry, which intuitively is a
consequence of the Mexican-hat form for the (pseudo)scalar potential. The
quantity $\phi_{N}$ corresponds then to the v.e.v. of $\sqrt{1/2}(\bar
{u}u+\bar{d}d),$ while $\phi_{S}$ to the v.e.v. of $\bar{s}s.$ Quite
interestingly, the term proportional to $h_{3}^{nyb}$ turns out to be
particularly important, since -together with the scalar condensates $\phi_{N}$
and $\phi_{S}$- it generates a mass difference between the $1^{-+}$ and
$1^{+-}$ hybrid nonets upon shifting the masses of the latter upwards (see
below). Note, the second term in Eq. (\ref{lelsmhybquad}) models the direct
contribution of the nonzero bare quark masses
\begin{equation}
\Delta^{hyb}=diag\{\delta_{N}^{hyb},\delta_{N}^{hyb},\delta_{S}^{hyb}\}
\end{equation}
and breaks flavor symmetry explicitly when $\delta_{S}^{hyb}\neq\delta
_{N}^{hyb}.$

The masses of the $1^{-+}$ exotic hybrid mesons read:
\begin{align}
m_{\pi_{1}^{hyb}}^{2}  &  =m_{1}^{hyb,2}+\frac{1}{2}(h_{1}^{hyb}+h_{2}%
^{hyb}+h_{3}^{hyb})\phi_{N}^{2}+\frac{h_{1}^{hyb}}{2}\phi_{S}^{2}+2\delta
_{N}^{hyb}\;,\label{mPi1}\\
m_{K_{1}^{hyb}}^{2}  &  =m_{1}^{hyb,2}+\frac{1}{4}\left(  2h_{1}^{hyb}%
+h_{2}^{hyb}\right)  \phi_{N}^{2}+\frac{1}{\sqrt{2}}\phi_{N}\phi_{S}%
h_{3}^{hyb}+\frac{1}{2}(h_{1}^{hyb}+h_{2}^{hyb})\phi_{S}^{2}+\delta_{N}%
^{hyb}+\delta_{S}^{hyb}\;,\label{mK1}\\
m_{\eta_{1,N}^{hyb}}^{2}  &  =m_{\pi_{1}}^{2}\;,\\
m_{\eta_{1,S}^{hyb}}^{2}  &  =m_{1}^{hyb,2}+\frac{h_{1}^{hyb}}{2}\phi_{N}%
^{2}+\left(  \frac{h_{1}^{hyb}}{2}+h_{2}^{hyb}+h_{3}^{hyb}\right)  \phi
_{S}^{2}+2\delta_{S}^{hyb}\;,
\end{align}
while the squared masses of the crypto-exotic pseudovector hybrid states are:
\begin{align}
m_{b_{1}^{hyb}}^{2}  &  =m_{1}^{hyb,2}+\frac{1}{2}(h_{1}^{hyb}+h_{2}%
^{hyb}-h_{3}^{hyb})\phi_{N}^{2}+\frac{h_{1}^{hyb}}{2}\phi_{S}^{2}+2\delta
_{N}^{hyb}\;,\\
m_{K_{1,B}^{hyb}}^{2}  &  =m_{1}^{hyb,2}+\frac{1}{4}\left(  2h_{1}^{hyb}%
+h_{2}^{hyb}\right)  \phi_{N}^{2}-\frac{1}{\sqrt{2}}\phi_{N}\phi_{S}%
h_{3}^{hyb}+\frac{1}{2}\left(  h_{1}^{hyb}+h_{2}^{hyb}\right)  \phi_{S}%
^{2}+\delta_{N}^{hyb}+\delta_{S}^{hyb}\;,\\
m_{h_{1N}^{hyb}}^{2}  &  =m_{b_{1}^{hyb}}^{2}\;,\\
m_{h_{1S}^{hyb}}^{2}  &  =m_{1}^{hyb2}+\frac{h_{1}^{hyb}}{2}\phi_{N}%
^{2}+\left(  \frac{h_{1}^{hyb}}{2}+h_{2}^{hyb}-h_{3}^{hyb}\right)  \phi
_{S}^{2}+2\delta_{S}^{hyb}\;. \label{mh1s}%
\end{align}
Note, these equations are formally equal to the mass expressions for vector
and axial-vector fields reported in Ref. \cite{Parganlija:2012fy} upon
replacing $h_{k}\rightarrow h_{k}^{hyb}$, $\delta_{k}\rightarrow\delta
_{k}^{hyb},$ and $m_{1}\rightarrow m_{1}^{hyb};$ this is expected, since the
terms are built following the same rules.\ There is however an important
difference: there is no $g_{1}^{hyb},$ since such a term is not possible for
the hybrid multiplet, see Appendix \ref{appB}.

In particular, we get the (exact) relations:
\begin{align}
m_{b_{1}^{hyb}}^{2}-m_{\pi_{1}^{hyb}}^{2}  &  =-2h_{3}^{hyb}\phi_{N}%
^{2}\,,\label{hcp1}\\
m_{K_{1,B}^{hyb}}^{2}-m_{K_{1}^{hyb}}^{2}  &  =-\sqrt{2}\phi_{N}\phi_{S}%
h_{3}^{hyb}\text{ ,}\label{hcp2}\\
m_{h_{1S}^{hyb}}^{2}-m_{\eta_{1,S}^{hyb}}^{2}  &  =-h_{3}^{hyb}\phi_{S}%
^{2}\,\text{.} \label{hcp3}%
\end{align}
Hence, only the parameter $h_{3}^{hyb}$ is responsible for the mass splitting
of the hybrid chiral partners.

\bigskip

Altogether, six parameters appear in the expressions for the hybrid masses,
but some simplifications are possible:\newline

a) The parameters $h_{1}^{hyb}$ and $m_{1}^{hyb}$ are not independent since
they always appear in the combination $m_{1}^{hyb,2}+\frac{1}{2}h_{1}%
^{hyb}(\phi_{N}^{2}+\phi_{S}^{2}).$ Hence, without loss of generality, we can
set $h_{1}^{hyb}=0.$ (In addition, the parameter $h_{1}^{hyb}\propto
N_{c}^{-2}$ is large-$N_{c}$ suppressed.) \newline

b) Only the difference $\delta_{S}^{hyb}-\delta_{N}^{hyb}$ is physical. In
fact, one can write
\begin{equation}
\mathrm{Tr}\left[  \Delta^{hyb}\left(  L_{\mu}^{hyb,2}+R_{\mu}^{hyb,2}\right)
\right]  =\mathrm{Tr}\left[  \left(  \Delta^{hyb}-\delta_{N}^{hyb}%
1_{3}\right)  \left(  L_{\mu}^{hyb,2}+R_{\mu}^{hyb,2}\right)  \right]
+\mathrm{Tr}\left[  \delta_{N}^{hyb}\left(  L_{\mu}^{hyb,2}+R_{\mu}%
^{hyb,2}\right)  \right]
\end{equation}
and the last term can be absorbed into the one proportional to $m_{1}^{hyb,2}$
(when $G$ is set equal to the condensate $G_{0}$). Therefore, for what
concerns masses, we set $\delta_{N}^{hyb}=0$. Moreover, considering that
\begin{equation}
\phi_{S}^{2}-2\phi_{N}^{2}\simeq0
\end{equation}
(this equation is exact in the $U(3)_{V}$ limit), one can neglect the
corresponding combinations in the expressions for the masses. As a result, the
parameter $h_{2}^{hyb}$ no longer appears and we are left with three
independent parameters
\begin{equation}
m_{\pi_{1}^{hyb}}^{2}\text{ , }h_{3}^{hyb},\text{ }\delta_{S}^{hyb}\,.
\end{equation}
We then obtain the following simple equations for the masses of the hybrid
states:
\begin{align}
m_{K_{1}^{hyb}}^{2}  &  \simeq m_{\pi_{1}^{hyb}}^{2}+\delta_{S}^{hyb}\text{
,}\\
m_{\eta_{1,N}^{hyb}}^{2}  &  \simeq m_{\pi_{1}^{hyb}}^{2}\;\text{,}\\
m_{\eta_{1,S}^{hyb}}^{2}  &  \simeq m_{\pi_{1}^{hyb}}^{2}+2\delta_{S}%
^{hyb}\text{ ,}\\
m_{b_{1}^{hyb}}^{2}  &  \simeq m_{\pi_{1}^{hyb}}^{2}-2h_{3}^{hyb}\phi_{N}%
^{2}\text{ ,}\\
m_{K_{1,B}^{hyb}}^{2}  &  \simeq m_{K_{1}^{hyb}}^{2}-\sqrt{2}\phi_{N}\phi
_{S}h_{3}^{hyb}\text{ ,}\\
m_{h_{1S}^{hyb}}^{2}  &  \simeq m_{\eta_{1,S}^{hyb}}^{2}-h_{3}^{hyb}\phi
_{S}^{2}\text{ .}%
\end{align}
Since the $s$-quark contribution is solely related to the strange constituent
quark, we shall use the numerical value obtained in the fit of Ref.
\cite{Parganlija:2012fy}
\begin{equation}
\delta_{S}^{hyb}\simeq\delta_{S}=0.151\text{ GeV}^{2}\,,
\end{equation}
which leaves us with two parameters that are fixed in the next section.

\subsection{Linear terms in the hybrid fields: hybrid decays}

The Lagrangian terms which generate decays of the hybrid states into
pseudovector and excited vector states as well as into (axial-)vector and
(pseudo)scalar mesons are given by:
\begin{align}
\mathcal{L}_{eLSM}^{\text{ hybrid-linear}}=  &  i\lambda_{1}^{hyb}%
G\mathrm{Tr}\left[  L_{\mu}^{hyb}(\tilde{\Phi}^{\mu}\Phi^{\dagger}-\Phi
\tilde{\Phi}^{\dag\mu})+R_{\mu}^{hyb}(\tilde{\Phi}^{\mu\dag}\Phi-\Phi^{\dag
}\tilde{\Phi}^{\mu})\,\right] \nonumber\\
&  +i\lambda_{2}^{hyb}\mathrm{Tr}([L_{\mu}^{hyb},L^{\mu}]\Phi\Phi^{\dagger
}+[R_{\mu}^{hyb},R^{\mu}]\Phi^{\dagger}\Phi)\nonumber\\
&  +\alpha^{hyb}\mathrm{Tr}(\tilde{L}_{\mu\nu}^{hyb}\Phi R^{\mu\nu}%
\Phi^{\dagger}-\tilde{R}_{\mu\nu}^{hyb}\Phi^{\dagger}L^{\mu\nu}\Phi)\text{
}\nonumber\\
&  +\beta_{A}^{hyb}(\det\Phi-\det\Phi^{\dag})\mathrm{Tr}(L_{\mu}%
^{hyb}(\partial^{\mu}\Phi\cdot\Phi^{\dag}-\Phi\cdot\partial^{\mu}\Phi^{\dag
})-R_{\mu}^{hyb}(\partial^{\mu}\Phi^{\dag}\cdot\Phi-\Phi^{\dag}\cdot
\partial^{\mu}\Phi))\text{ .} \label{hybriddecay}%
\end{align}
These terms are invariant under $SU(3$)$_{R}\times SU(3)_{L}$, $C,$ and $P$
transformations. The first three terms are invariant under $U(3$)$_{R}\times
U(3)_{L}$, while the last breaks $U_{A}(1)$: this is a typical term caused by
the axial anomaly \cite{Giacosa:2017pos}. In addition, the first two terms are
also dilatation invariant: the two coupling constants $\lambda_{1}^{hyb}$ and
$\lambda_{2}^{hyb}$ are dimensionless. The third term, proportional to
$\alpha^{hyb}$, involves the Levi-Civita tensor and carries the dimension
Energy$^{-2}$, while the fourth $\beta_{A}^{hyb}$ has dimension Energy$^{-3}.$
In the Appendix \ref{appC} we report the proof of the invariance properties
for each of these terms.

Let us now consider the first term closely. Upon condensation of the glueball
field $G$, the effective coupling $\lambda_{1}^{hyb}G_{0}$ has dimension
energy. In terms of the physical nonets, the first term reads
\begin{align}
\mathcal{L}_{eLSM,1}^{\text{ hybrid-linear}}=  &  i2\lambda_{1}^{hyb}G\left\{
\mathrm{Tr}\left[  \Pi_{\mu}^{hyb}\,[P,B^{\mu}]\right]  +\mathrm{Tr}\left[
\Pi_{\mu}^{hyb}\,[V_{E}^{\mu},S]\right]  \right\} \\
&  +2\lambda_{1}^{hyb}G\left\{  \mathrm{Tr}\left[  B_{\mu}^{hyb}\,\left\{
P,V_{E}^{\mu}\right\}  \right]  +\mathrm{Tr}\left[  B_{\mu}^{hyb}\,\left\{
B^{\mu},S\right\}  \right]  \right\}  .
\end{align}
It generates decays of the type $\Pi^{hyb}\rightarrow BP$ , in particular:
\begin{equation}
\pi_{1}\rightarrow b_{1}(1230)\pi\text{ .}%
\end{equation}
These decay channels of exotic hybrids are expected to be dominant.
Correspondingly, also the decay $B^{hyb}\rightarrow V_{E}P$ takes place. Note
that further decays of the form $\Pi^{hyb}\rightarrow V_{E}S$ and
$B^{hyb}\rightarrow\,B^{\mu}S$ cannot take place because they are
kinematically forbidden.

We now turn to the second term. When the matrix $\Phi$ condenses, $\Phi
\Phi^{\dagger}=\Phi_{0}^{2}$, this term vanishes: there is no mixing between
(axial-)vector mesons and vectorial hybrid states, in agreement with the fact
that they have different $C$-parity. A related important consideration is the
lack of a term that generates a mixing of the hybrid states with
(pseudo)scalar mesons, see Appendix \ref{appB} for details. As a consequence,
no shift of the hybrid fields and no additional renormalization factor for
(pseudo)scalar states needs to be performed. The necessary shifts are those of
the \textquotedblleft standard eLSM\textquotedblright\ that were studied in
Ref.~\cite{Parganlija:2012fy} and are summarized in Appendix \ref{appA}. The
second term can be cast into the form:%
\begin{equation}
\mathcal{L}_{eLSM,2}^{\text{ hybrid-linear}}=2i\lambda_{2}^{hyb}%
\mathrm{Tr}\left[  \left(  [\Pi_{\mu}^{hyb},V^{\mu}]+[B_{\mu}^{hyb},A^{\mu
}]\right)  \left(  S^{2}+P^{2}\right)  \right]  -2\lambda_{2}^{hyb}%
\mathrm{Tr}\left[  \left(  [\Pi_{\mu}^{hyb},A^{\mu}]+[B_{\mu}^{hyb},V^{\mu
}]\right)  [P,S]\right]  \text{ .}%
\end{equation}
Thus, we get decays of the types $\Pi^{hyb}\rightarrow VPP$ and $\Pi
^{hyb}\rightarrow A^{\mu}PS$. The decay channel into $VPP\ $is potentially
relevant. For the nonet $B_{\mu}^{hyb},$ decays into $A^{\mu}PP$ are expected.
As a next step, one has to perform the transformations described in Appendix
\ref{appA} (shifts of $S$ and $A^{\mu}$ and redefinition of $P$), and other
decays emerge, such as the one into two pseudoscalar states. The decays
$\pi_{1}\rightarrow\eta\pi$ and $\pi_{1}\rightarrow\eta^{\prime}\pi$, however,
do not follow from this term.

The third term in Eq.(\ref{hybriddecay}) breaks dilation invariance but leads
to two interesting decay channels: $\Pi^{hyb}\rightarrow VP$ and
$B^{hyb}\rightarrow AP$. In fact, the most relevant decay terms read:%
\begin{equation}
\mathcal{L}_{eLSM,3}^{\text{ hybrid-linear}}=i\alpha^{hyb}\phi_{N}\left\{
\mathrm{Tr}(\tilde{\Pi}_{\mu\nu}^{hyb}[P,V^{\mu\nu}])\,-\mathrm{Tr}(\tilde
{B}_{\mu\nu}^{hyb}([P,A^{\mu\nu}])\,\right\}  +...\text{ ,}%
\end{equation}
where $\phi_{N}$ is the condensate of $\sigma_{N}.$ Hence, this term is is
responsible for $\pi_{1}\rightarrow\rho\pi$. This is the channel in which
$\pi_{1}(1600)$ was recently observed at COMPASS \cite{Akhunzyanov:2018lqa}.
Terms that make use of the Levi-Civita tensor (here into $\tilde{L}_{\mu\nu
}^{hyb}=\frac{1}{2}\varepsilon_{\mu\nu\rho\sigma}L^{hyb,\rho\sigma}$ and
$\tilde{R}_{\mu\nu}^{hyb}=\frac{1}{2}\varepsilon_{\mu\nu\rho\sigma}%
R^{hyb,\rho\sigma}$) are linked to the axial anomaly and are typically not
negligible, even if the corresponding coupling constant is not dimensionless.

As a last step, we consider the fourth term in Eq.(\ref{hybriddecay}). This
term breaks explicitly the $U_{A}(1)$ symmetry because of the involvement of
the determinant. Considering that $\det\Phi-\det\Phi^{\dag}=i\frac{Z_{\pi}}%
{2}\sqrt{\frac{3}{2}}\phi_{N}^{2}\eta_{0}+...$ \cite{Olbrich:2017fsd}, one
has:
\begin{equation}
\mathcal{L}_{eLSM,4}^{\text{ hybrid-linear}}=-\beta_{A}^{hyb}Z_{\pi}%
\sqrt{\frac{3}{2}}\phi_{N}^{3}\eta_{0}\mathrm{Tr}(\Pi_{\mu}^{hyb}\partial
^{\mu}P)+...\text{,}%
\end{equation}
hence decays of the type $\Pi^{hyb}\rightarrow P\eta_{0}$ emerge. Since
$\eta_{0}$ is a combination of $\eta$ and $\eta^{\prime},$ the decays $\pi
_{1}\rightarrow\pi\eta$ and $\pi_{1}\rightarrow\pi\eta^{\prime}$ follow. Note,
experimentally the decay $\pi_{1}(1400)\rightarrow\pi\eta$ and the decay
$\pi_{1}(1600)\rightarrow\pi\eta^{\prime}$ have been seen in experiments. If
these resonances ultimately correspond to a unique hybrid state
\cite{Rodas:2018owy}, it means that both decay channels have been measured.
Similar decay terms appear for the other members of the nonet. It is
interesting to notice that this fourth term does not lead to two-body decays
for the nonet of chiral partners $B^{hyb}.$

As a last remark, we recall that the structure $\det\Phi-\det\Phi^{\dag}$
mixes with the pseudoscalar glueball \cite{Eshraim:2012jv,Eshraim:2016mds}.
Hence, the following interaction term is possible:
\begin{align}
\mathcal{L}_{eLSM,\tilde{G}}^{\text{ hybrid-linear}}  &  =i\beta_{\tilde{G}%
}^{hyb}\tilde{G}\mathrm{Tr}(L_{\mu}^{hyb}(\partial^{\mu}\Phi\cdot\Phi^{\dag
}-\Phi\cdot\partial^{\mu}\Phi^{\dag})-R_{\mu}^{hyb}(\partial^{\mu}\Phi^{\dag
}\cdot\Phi-\Phi^{\dag}\cdot\partial^{\mu}\Phi))\text{ }\nonumber\\
&  =-2\beta_{\tilde{G}}^{hyb}\tilde{G}\phi_{N}\mathrm{Tr}(\Pi_{\mu}%
^{hyb}\partial^{\mu}P)+...
\end{align}
An interesting consequence is the decay $\tilde{G}\rightarrow\pi_{1}^{hyb}\pi
$. According to lattice QCD, the mass of the pseudoscalar glueball $\tilde{G}$
may be in the range around $2.6$ GeV \cite{Chen:2005mg}, therefore this decay
is kinematically allowed. The detailed study of this term is left for the
future, when the pseudoscalar glueball will be supported by concrete
experimental candidates.

\begin{table}[h]
\centering%
\begin{tabular}
[t]{|c|c|}\hline
Resonance & Mass [MeV] \rule{0pt}{3.7ex}\rule[-2ex]{0pt}{0pt}\\\hline
$\pi_{1}^{hyb}$ & 1660 \rule{0pt}{3ex}\rule[-1ex]{0pt}{0pt}[\textbf{input}
using $\pi_{1}(1600)$ \cite{Tanabashi:2018oca}]\\\hline
$\eta_{1,N}^{hyb}$ & 1660\\\hline
$\eta_{1,S}^{hyb}$ & 1751 \rule{0pt}{3ex}\rule[-1ex]{0pt}{0pt}\\\hline
$K_{1}^{hyb}$ & 1707\\\hline
$b_{1}^{hyb}$ & 2000 \rule{0pt}{3ex}\rule[-1ex]{0pt}{0pt}[\textbf{input} set
as an estimate]\\\hline
$h_{1N,B}^{hyb}$ & 2000 \rule{0pt}{3ex}\rule[-1ex]{0pt}{0pt}\\\hline
$K_{1,B}^{hyb}$ & 2063\\\hline
$h_{1S,B}^{hyb}$ & 2126\\\hline
\end{tabular}
\hspace*{2cm}
\begin{tabular}
[t]{|c|c|}\hline
Ratio & Value \rule{0pt}{3.7ex}\rule[-2ex]{0pt}{0pt}\\\hline
$\Gamma_{K_{1}^{hyb}\rightarrow Kh_{1}(1170)}/\Gamma_{\pi_{1}^{hyb}%
\rightarrow\pi b_{1}}$ & 0.050 \rule{0pt}{3.7ex}\rule[-2ex]{0pt}{0pt}\\\hline
$\Gamma_{b_{1}^{hyb}\rightarrow\pi\omega(1650)}/\Gamma_{\pi_{1}^{hyb}%
\rightarrow\pi b_{1}}$ & 0.065 \rule{0pt}{3.7ex}\rule[-2ex]{0pt}{0pt}\\\hline
$\Gamma_{K_{1B}^{hyb}\rightarrow\pi K^{\ast}(1680)}/\Gamma_{\pi_{1}%
^{hyb}\rightarrow\pi b_{1}}$ & 0.19 \rule{0pt}{3.7ex}\rule[-2ex]%
{0pt}{0pt}\\\hline
$\Gamma_{h_{1,N}^{hyb}\rightarrow\pi\rho(1700)}/\Gamma_{\pi_{1}^{hyb}%
\rightarrow\pi b_{1}}$ & 0.16 \rule{0pt}{3.7ex}\rule[-2ex]{0pt}{0pt}\\\hline
\end{tabular}
\caption{\textit{Left:} Masses of the exotic $J^{PC}=1^{-+}$ and
$J^{PC}=1^{+-}$ hybrid mesons. \textit{Right:} Ratios for the decay of vector
and pseudovector hybrid mesons into pseudoscalar, pseudovector, and excited
vector mesons (term proportional to $\lambda_{1}^{\text{hyb}}$ in Eq.
(\ref{hybriddecay})).}%
\label{tabIII-IV}%
\end{table}

\begin{table}[t]%
\begin{tabular}
[t]{|c|c|}\hline
Ratio & Value\rule{0pt}{3ex}\rule[-2ex]{0pt}{0pt}\\\hline
$\Gamma_{K_{1}^{0hyb}\rightarrow K^{-}\pi^{+}}/\Gamma_{b_{1}^{0hyb}%
\rightarrow\pi^{+}\pi^{-}\eta}$ & $0.0041$ \rule{0pt}{3ex}\rule[-2ex]%
{0pt}{0pt}\\\hline
$\Gamma_{K_{1}^{0hyb}\rightarrow\overline{K}^{0}\eta}/\Gamma_{b_{1}%
^{0hyb}\rightarrow\pi^{+}\pi^{-}\eta}$ & $0.0027$ \rule{0pt}{3ex}%
\rule[-2ex]{0pt}{0pt}\\\hline
$\Gamma_{K_{1}^{0hyb}\rightarrow\overline{K}^{0}\eta^{\prime}}/\Gamma
_{b_{1}^{0hyb}\rightarrow\pi^{+}\pi^{-}\eta}$ & $3.6\cdot10^{-7}$
\rule{0pt}{3ex}\rule[-2ex]{0pt}{0pt}\\\hline
$\Gamma_{b_{1}^{0hyb}\rightarrow\pi^{+}a_{0}^{-}}/\Gamma_{b_{1}^{0hyb}%
\rightarrow\pi^{+}\pi^{-}\eta}$ & $0.24$ \rule{0pt}{3ex}\rule[-2ex]%
{0pt}{0pt}\\\hline
$\Gamma_{b_{1}^{0hyb}\rightarrow K^{+}K_{S}^{-}}/\Gamma_{b_{1}^{0hyb}%
\rightarrow\pi^{+}\pi^{-}\eta}$ & $0.0083$\rule{0pt}{3ex}\rule[-2ex]%
{0pt}{0pt}\\\hline
$\Gamma_{b_{1}^{+hyb}\rightarrow K^{+}\overline{K}^{\ast0}}/\Gamma
_{b_{1}^{0hyb}\rightarrow\pi^{+}\pi^{-}\eta}$ & $0.0011$ \rule{0pt}{3ex}%
\rule[-2ex]{0pt}{0pt}\\\hline
$\Gamma_{h_{1N,B}^{hyb}\rightarrow K^{0}K_{S}^{0}}/\Gamma_{b_{1}%
^{0hyb}\rightarrow\pi^{+}\pi^{-}\eta}$ & $0.0082$ \rule{0pt}{3ex}%
\rule[-2ex]{0pt}{0pt}\\\hline
$\Gamma_{h_{1N,B}^{hyb}\rightarrow K^{0}\overline{K}^{\ast0}}/\Gamma
_{b_{1}^{0hyb}\rightarrow\pi^{+}\pi^{-}\eta}$ & $0.0015$ \rule{0pt}{3ex}%
\rule[-2ex]{0pt}{0pt}\\\hline
$\Gamma_{h_{1S,B}^{hyb}\rightarrow K^{0}\overline{K}_{S}^{0}}/\Gamma
_{b_{1}^{0hyb}\rightarrow\pi^{+}\pi^{-}\eta}$ & $0.11$ \rule{0pt}{3ex}%
\rule[-2ex]{0pt}{0pt}\\\hline
$\Gamma_{h_{1S,B}^{hyb}\rightarrow K^{0}\overline{K}^{\ast0}}/\Gamma
_{b_{1}^{0hyb}\rightarrow\pi^{+}\pi^{-}\eta}$ & $0.0031$ \rule{0pt}{3ex}%
\rule[-2ex]{0pt}{0pt}\\\hline
$\Gamma_{K_{1,B}^{0hyb}\rightarrow\overline{K}^{0}\sigma_{1}}\Gamma
_{b_{1}^{0hyb}\rightarrow\pi^{+}\pi^{-}\eta}$ & $0.034$ \rule{0pt}{3ex}%
\rule[-2ex]{0pt}{0pt}\\\hline
$\Gamma_{K_{1,B}^{0hyb}\rightarrow\overline{K}^{0}\omega_{N}}/\Gamma
_{b_{1}^{0hyb}\rightarrow\pi^{+}\pi^{-}\eta}$ & $0.0072$ \rule{0pt}{3ex}%
\rule[-2ex]{0pt}{0pt}\\\hline
$\Gamma_{K_{1,B}^{0hyb}\rightarrow\overline{K}^{0}\omega_{S}}/\Gamma
_{b_{1}^{0hyb}\rightarrow\pi^{+}\pi^{-}\eta}$ & $0.0098$ \rule{0pt}{3ex}%
\rule[-2ex]{0pt}{0pt}\\\hline
$\Gamma_{K_{1,B}^{0hyb}\rightarrow\overline{K}_{S}^{0}\pi^{0}}/\Gamma
_{b_{1}^{0hyb}\rightarrow\pi^{+}\pi^{-}\eta}$ & $0.10$ \rule{0pt}{3ex}%
\rule[-2ex]{0pt}{0pt}\\\hline
$\Gamma_{\overline{K}_{1,B}^{0hyb}\rightarrow K^{0}\rho^{0}}/\Gamma
_{b_{1}^{0hyb}\rightarrow\pi^{+}\pi^{-}\eta}$ & $0.28$ \rule{0pt}{3ex}%
\rule[-2ex]{0pt}{0pt}\\\hline
$\Gamma_{\overline{K}_{1,B}^{0hyb}\rightarrow K_{S}^{0}\eta}/\Gamma
_{b_{1}^{0hyb}\rightarrow\pi^{+}\pi^{-}\eta}$ & $0.030$ \rule{0pt}{3ex}%
\rule[-2ex]{0pt}{0pt}\\\hline
$\Gamma_{\overline{K}_{1,B}^{0hyb}\rightarrow K^{0}a_{0}^{0}}/\Gamma
_{b_{1}^{0hyb}\rightarrow\pi^{+}\pi^{-}\eta}$ & $0.0092$ \rule{0pt}{3ex}%
\rule[-2ex]{0pt}{0pt}\\\hline
\end{tabular}
\hspace*{2cm}
\begin{tabular}
[t]{|c|c|}\hline
Ratio & Value\rule{0pt}{3ex}\rule[-2ex]{0pt}{0pt}\\\hline
$\Gamma_{\pi_{1}^{0hyb}\rightarrow K^{\ast0}\overline{K}^{0}\pi^{0}}%
/\Gamma_{b_{1}^{0hyb}\rightarrow\pi^{+}\pi^{-}\eta}$ & $0.0046$
\rule{0pt}{3ex}\rule[-2ex]{0pt}{0pt}\\\hline
$\Gamma_{\pi_{1}^{+hyb}\rightarrow\pi^{0}\rho^{+}\eta}/\Gamma_{b_{1}%
^{0hyb}\rightarrow\pi^{+}\pi^{-}\eta}$ & $0.1832$ \rule{0pt}{3ex}
\rule[-2ex]{0pt}{0pt}\\\hline
$\Gamma_{\eta_{1N}^{hyb}\rightarrow K^{\ast0}\overline{K}^{0}\pi^{0}}%
/\Gamma_{b_{1}^{0hyb}\rightarrow\pi^{+}\pi^{-}\eta}$ & $0.0046$
\rule{0pt}{3ex}\rule[-2ex]{0pt}{0pt}\\\hline
$\Gamma_{\eta_{1S}^{hyb}\rightarrow K^{\ast0}\overline{K}^{0}\pi^{0}}%
/\Gamma_{b_{1}^{0hyb}\rightarrow\pi^{+}\pi^{-}\eta}$ & $0.024$\rule{0pt}{3ex}%
\rule[-2ex]{0pt}{0pt}\\\hline
$\Gamma_{K_{1}^{0hyb}\rightarrow\overline{K}^{0}\pi^{0}\rho^{0}}/\Gamma
_{b_{1}^{0hyb}\rightarrow\pi^{+}\pi^{-}\eta}$ & $0.022$ \rule{0pt}{3ex}
\rule[-2ex]{0pt}{0pt}\\\hline
$\Gamma_{K_{1}^{0hyb}\rightarrow\overline{K}^{0}\pi^{0}\omega_{N}}%
/\Gamma_{b_{1}^{0hyb}\rightarrow\pi^{+}\pi^{-}\eta}$ & $0.021$ \rule{0pt}{3ex}%
\rule[-2ex]{0pt}{0pt}\\\hline
$\Gamma_{K_{1}^{0hyb}\rightarrow\overline{K}^{0}\pi^{0}\omega_{S}}%
/\Gamma_{b_{1}^{0hyb}\rightarrow\pi^{+}\pi^{-}\eta}$ & $0.0012$
\rule{0pt}{3ex}\rule[-2ex]{0pt}{0pt}\\\hline
$\Gamma_{K_{1}^{0hyb}\rightarrow\overline{K}^{0}\pi^{0}\pi^{0}}/\Gamma
_{b_{1}^{0hyb}\rightarrow\pi^{+}\pi^{-}\eta}$ & $0.14$ \rule{0pt}{3ex}
\rule[-2ex]{0pt}{0pt}\\\hline
$\Gamma_{K_{1}^{0hyb}\rightarrow\overline{K}^{0}\pi^{0}\eta}/\Gamma
_{b_{1}^{0hyb}\rightarrow\pi^{+}\pi^{-}\eta}$ & $0.0090$ \rule{0pt}{3ex}
\rule[-2ex]{0pt}{0pt}\\\hline
$\Gamma_{b_{1}^{0hyb}\rightarrow K^{0}\overline{K}^{0}\pi^{0}}/\Gamma
_{b_{1}^{0hyb}\rightarrow\pi^{+}\pi^{-}\eta}$ & $0.17$\rule{0pt}{3ex}
\rule[-2ex]{0pt}{0pt}\\\hline
$\Gamma_{b_{1}^{0hyb}\rightarrow\pi^{+}\pi^{-}\eta^{\prime}}/\Gamma
_{b_{1}^{0hyb}\rightarrow\pi^{+}\pi^{-}\eta}$ & $0.59$\rule{0pt}{3ex}
\rule[-2ex]{0pt}{0pt}\\\hline
$\Gamma_{b_{1}^{0hyb}\rightarrow a_{1}^{-}\pi^{+}\eta}/\Gamma_{b_{1}%
^{0hyb}\rightarrow\pi^{+}\pi^{-}\eta}$ & $0.015$\rule{0pt}{3ex} \rule[-2ex]%
{0pt}{0pt}\\\hline
$\Gamma_{b_{1}^{0hyb}\rightarrow K^{-}K^{+}\eta}/\Gamma_{b_{1}^{0hyb}%
\rightarrow\pi^{+}\pi^{-}\eta}$ & $0.00031$\rule{0pt}{3ex}\rule[-2ex]%
{0pt}{0pt}\\\hline
$\Gamma_{b_{1}^{0hyb}\rightarrow K^{-}K^{+}\eta^{\prime}}/\Gamma_{b_{1}%
^{0hyb}\rightarrow\pi^{+}\pi^{-}\eta}$ & $0.000034$ \rule{0pt}{3ex}
\rule[-2ex]{0pt}{0pt}\\\hline
$\Gamma_{b_{1}^{0hyb}\rightarrow K^{+}\overline{K}_{1}^{0}\pi^{-}}%
/\Gamma_{b_{1}^{0hyb}\rightarrow\pi^{+}\pi^{-}\eta}$ & $0.0029$
\rule{0pt}{3ex}\rule[-2ex]{0pt}{0pt}\\\hline
$\Gamma_{h_{1N,B}^{hyb}\rightarrow K^{-}K_{1}^{+}\pi^{0}}/\Gamma_{b_{1}%
^{0hyb}\rightarrow\pi^{+}\pi^{-}\eta}$ & $0.0016$ \rule{0pt}{3ex}
\rule[-2ex]{0pt}{0pt}\\\hline
$\Gamma_{h_{1N,B}^{hyb}\rightarrow K^{0}\overline{K}^{0}\eta}/\Gamma
_{b_{1}^{0hyb}\rightarrow\pi^{+}\pi^{-}\eta}$ & $0.00092$ \rule{0pt}{3ex}
\rule[-2ex]{0pt}{0pt}\\\hline
$\Gamma_{h_{1N,B}^{hyb}\rightarrow K^{0}\overline{K}^{0}\eta^{\prime}}%
/\Gamma_{b_{1}^{0hyb}\rightarrow\pi^{+}\pi^{-}\eta}$ & $1.7\times10^{-6}$
\rule{0pt}{3ex}\rule[-2ex]{0pt}{0pt}\\\hline
$\Gamma_{h_{1N,B}^{hyb}\rightarrow K^{0}\overline{K}^{0}\pi^{0}}/\Gamma
_{b_{1}^{0hyb}\rightarrow\pi^{+}\pi^{-}\eta}$ & $0.17$\rule{0pt}{3ex}
\rule[-2ex]{0pt}{0pt}\\\hline
$\Gamma_{h_{1S,B}^{hyb}\rightarrow K_{1}^{+}K^{-}\pi^{0}}/\Gamma_{b_{1}%
^{0hyb}\rightarrow\pi^{+}\pi^{-}\eta}$ & $0.016$ \rule[-2ex]{0pt}{0pt}\\\hline
$\Gamma_{h_{1S,B}^{hyb}\rightarrow K^{0}\overline{K}^{0}\eta^{\prime}}%
/\Gamma_{b_{1}^{0hyb}\rightarrow\pi^{+}\pi^{-}\eta}$ & $0.00012$
\rule{0pt}{3ex}\rule[-2ex]{0pt}{0pt}\\\hline
$\Gamma_{h_{1S,B}^{hyb}\rightarrow K^{0}\overline{K}^{0}\eta}/\Gamma
_{b_{1}^{0hyb}\rightarrow\pi^{+}\pi^{-}\eta}$ & $0.0036$ \rule{0pt}{3ex}
\rule[-2ex]{0pt}{0pt}\\\hline
$\Gamma_{h_{1S,B}^{hyb}\rightarrow K^{0}\overline{K}^{0}\pi^{0}}/\Gamma
_{b_{1}^{0hyb}\rightarrow\pi^{+}\pi^{-}\eta}$ & $0.48$\rule{0pt}{3ex}
\rule[-2ex]{0pt}{0pt}\\\hline
\end{tabular}
\caption{\textit{Left:} Ratios for the two-body decay of vector and
pseudovector hybrid mesons into axial-vector and pseudo(scalar) mesons(term
proportional to $\lambda_{2}^{\text{hyb}}$ in Eq. (\ref{hybriddecay}%
)).\newline\textit{Right:} Ratios for the three-body decay of vector and
pseudovector hybrid mesons into axial(vector) and pseudo-scalar mesons (term
proportional to $\lambda_{2}^{\text{hyb}}$ in Eq. (\ref{hybriddecay})).}%
\label{tabV-VI}%
\end{table}

\begin{table}[t]%
\begin{tabular}
[t]{|c|c|}\hline
Ratio & Value\rule{0pt}{3.7ex}\rule[-2ex]{0pt}{0pt}\\\hline
$\Gamma_{K_{1,B}^{0hyb}\rightarrow f_{1N}\overline{K}^{0}\pi^{0}}%
/\Gamma_{b_{1}^{0hyb}\rightarrow\pi^{+}\pi^{-}\eta}$ & $0.0012$
\rule{0pt}{3.7ex}\rule[-2ex]{0pt}{0pt}\\\hline
$\Gamma_{K_{1,B}^{0hyb}\rightarrow f_{1S}\overline{K}^{0}\pi^{0}}%
/\Gamma_{b_{1}^{0hyb}\rightarrow\pi^{+}\pi^{-}\eta}$ & $0.001$
\rule{0pt}{3.7ex}\rule[-2ex]{0pt}{0pt}\\\hline
$\Gamma_{K_{1,B}^{0hyb}\rightarrow\overline{K}_{1}^{0}\pi^{0}\eta}%
/\Gamma_{b_{1}^{0hyb}\rightarrow\pi^{+}\pi^{-}\eta}$ & $0.00062$
\rule{0pt}{3.7ex}\rule[-2ex]{0pt}{0pt}\\\hline
$\Gamma_{K_{1,B}^{0hyb}\rightarrow\overline{K}_{1}^{0}\pi^{0}\pi^{0}}%
/\Gamma_{b_{1}^{0hyb}\rightarrow\pi^{+}\pi^{-}\eta}$ & $0.066$%
\rule{0pt}{3.7ex}\rule[-2ex]{0pt}{0pt}\\\hline
$\Gamma_{\overline{K}_{1,B}^{0hyb}\rightarrow\overline{K}^{0}\pi^{0}a_{1}^{0}%
}/\Gamma_{b_{1}^{0hyb}\rightarrow\pi^{+}\pi^{-}\eta}$ & $0.0032$
\rule{0pt}{3.7ex}\rule[-2ex]{0pt}{0pt}\\\hline
$\Gamma_{\overline{K}_{1,B}^{0hyb}\rightarrow\overline{K}^{0}K^{-}K^{+}%
}/\Gamma_{b_{1}^{0hyb}\rightarrow\pi^{+}\pi^{-}\eta}$ & $0.074$
\rule{0pt}{3.7ex}\rule[-2ex]{0pt}{0pt}\\\hline
$\Gamma_{\overline{K}_{1,B}^{0hyb}\rightarrow\overline{K}^{0}\pi^{0}\eta
}/\Gamma_{b_{1}^{0hyb}\rightarrow\pi^{+}\pi^{-}\eta}$ & $0.12$
\rule{0pt}{3.7ex}\rule[-2ex]{0pt}{0pt}\\\hline
$\Gamma_{\overline{K}_{1,B}^{0hyb}\rightarrow\overline{K}^{0}\pi^{0}%
\eta^{\prime}}/\Gamma_{b_{1}^{0hyb}\rightarrow\pi^{+}\pi^{-}\eta}$ &
$0.000053$ \rule{0pt}{3.7ex}\rule[-2ex]{0pt}{0pt}\\\hline
$\Gamma_{\overline{K}_{1,B}^{0hyb}\rightarrow\overline{K}^{0}\eta\eta}%
/\Gamma_{b_{1}^{0hyb}\rightarrow\pi^{+}\pi^{-}\eta}$ & $0.0029$
\rule{0pt}{3.7ex}\rule[-2ex]{0pt}{0pt}\\\hline
\end{tabular}
\hspace*{2cm}
\begin{tabular}
[t]{|c|c|}\hline
Ratio & Value\rule{0pt}{3.7ex}\rule[-2ex]{0pt}{0pt}\\\hline
$\Gamma_{\pi_{1}^{0hyb}\rightarrow\overline{K}^{0}K^{\ast0}}/\Gamma_{\pi
_{1}^{-hyb}\rightarrow\rho^{0}\pi^{-}}$ & $0.61$ \rule{0pt}{3.7ex}
\rule[-2ex]{0pt}{0pt}\\\hline
$\Gamma_{\eta_{1N}^{hyb}\rightarrow\overline{K}^{0}K^{\ast0}}/\Gamma_{\pi
_{1}^{-hyb}\rightarrow\rho^{0}\pi^{-}}$ & $0.61$ \rule{0pt}{3.7ex}
\rule[-2ex]{0pt}{0pt}\\\hline
$\Gamma_{\eta_{1S}^{hyb}\rightarrow\overline{K}^{0}K^{\ast0}}/\Gamma_{\pi
_{1}^{-hyb}\rightarrow\rho^{0}\pi^{-}}$ & $1.6$ \rule{0pt}{3.7ex}
\rule[-2ex]{0pt}{0pt}\\\hline
$\Gamma_{K_{1}^{0hyb}\rightarrow K^{0}\omega_{S}}/\Gamma_{\pi_{1}%
^{-hyb}\rightarrow\rho^{0}\pi^{-}}$ & $0.00022$ \rule{0pt}{3.7ex}
\rule[-2ex]{0pt}{0pt}\\\hline
$\Gamma_{K_{1}^{0hyb}\rightarrow\overline{K}^{\ast0}\eta}/\Gamma_{\pi
_{1}^{-hyb}\rightarrow\rho^{0}\pi^{-}}$ & $0.0011$ \rule{0pt}{3.7ex}
\rule[-2ex]{0pt}{0pt}\\\hline
$\Gamma_{K_{1}^{0hyb}\rightarrow K^{\ast0}\pi^{0}}/\Gamma_{\pi_{1}%
^{-hyb}\rightarrow\rho^{0}\pi^{-}}$ & $0.00022$ \rule{0pt}{3.7ex}
\rule[-2ex]{0pt}{0pt}\\\hline
$\Gamma_{K_{1}^{0hyb}\rightarrow\overline{K}^{0}\rho^{0}}/\Gamma_{\pi
_{1}^{-hyb}\rightarrow\rho^{0}\pi^{-}}$ & $0.0011$ \rule{0pt}{3.7ex}
\rule[-2ex]{0pt}{0pt}\\\hline
$\Gamma_{K_{1}^{0hyb}\rightarrow K^{0}\omega_{N}}/\Gamma_{\pi_{1}%
^{-hyb}\rightarrow\rho^{0}\pi^{-}}$ & $0.0011$ \rule{0pt}{3.7ex}
\rule[-2ex]{0pt}{0pt}\\\hline
$\Gamma_{b_{1}^{0hyb}\rightarrow\pi^{-}a_{1}^{+}}/\Gamma_{\pi_{1}%
^{-hyb}\rightarrow\rho^{0}\pi^{-}}$ & $3.8$ \rule{0pt}{3.7ex}\rule[-2ex]%
{0pt}{0pt}\\\hline
$\Gamma_{b_{1}^{0hyb}\rightarrow\overline{K}_{1}^{0}K^{0}}/\Gamma_{\pi
_{1}^{-hyb}\rightarrow\rho^{0}\pi^{-}}$ & $0.60$ \rule{0pt}{3.7ex}
\rule[-2ex]{0pt}{0pt}\\\hline
$\Gamma_{h_{1N,B}^{hyb}\rightarrow\overline{K}_{1}^{0}K^{0}}/\Gamma_{\pi
_{1}^{-hyb}\rightarrow\rho^{0}\pi^{-}}$ & $0.59$ \rule{0pt}{3.7ex}
\rule[-2ex]{0pt}{0pt}\\\hline
$\Gamma_{h_{1S,B}^{hyb}\rightarrow\overline{K}_{1}^{0}K^{0}}/\Gamma_{\pi
_{1}^{-hyb}\rightarrow\rho^{0}\pi^{-}}$ & $1.7801$ \rule{0pt}{3.7ex}
\rule[-2ex]{0pt}{0pt}\\\hline
$\Gamma_{K_{1,B}^{0hyb}\rightarrow\overline{K}^{0}f_{1S}}/\Gamma_{\pi
_{1}^{-hyb}\rightarrow\rho^{0}\pi^{-}}$ & $0.60$ \rule{0pt}{3.7ex}
\rule[-2ex]{0pt}{0pt}\\\hline
$\Gamma_{K_{1,B}^{0hyb}\rightarrow\overline{K}^{0}f_{1N}}/\Gamma_{\pi
_{1}^{-hyb}\rightarrow\rho^{0}\pi^{-}}$ & $1.78$ \rule{0pt}{3.7ex}
\rule[-2ex]{0pt}{0pt}\\\hline
$\Gamma_{K_{1,B}^{0hyb}\rightarrow K_{1}^{0}\eta}/\Gamma_{\pi_{1}%
^{-hyb}\rightarrow\rho^{0}\pi^{-}}$ & $0.010$ \rule{0pt}{3.7ex}\rule[-2ex]%
{0pt}{0pt}\\\hline
$\Gamma_{K_{1,B}^{0hyb}\rightarrow\overline{K}_{1}^{0}\pi^{0}}/\Gamma_{\pi
_{1}^{-hyb}\rightarrow\rho^{0}\pi^{-}}$ & $0.029$ \rule{0pt}{3.7ex}
\rule[-2ex]{0pt}{0pt}\\\hline
$\Gamma_{K_{1,B}^{0hyb}\rightarrow\overline{K}^{0}a_{1}^{0}}/\Gamma_{\pi
_{1}^{-hyb}\rightarrow\rho^{0}\pi^{-}}$ & $0.046$ \rule{0pt}{3.7ex}
\rule[-2ex]{0pt}{0pt}\\\hline
\end{tabular}
\caption{\textit{Left:} Ratios for the decay of vector and pseudovector hybrid
mesons pseudoscalar, pseudovector, and excited vector mesons (term
proportional to $\lambda_{2}^{\text{hyb}}$ in Eq. (\ref{hybriddecay})).
\textit{Right:} Ratios for the decay of vector and pseudovector hybrid mesons
pseudoscalar, pseudovector, and excited vector mesons (term proportional to
$\alpha^{\text{hyb}}$ in Eq. (\ref{hybriddecay})).}%
\label{tabVII-VIII}%
\end{table}

\begin{table}[t]
\centering%
\begin{tabular}
[t]{|c|c|}\hline
Ratio & Value\\\hline
$\Gamma_{\pi_{1}^{hyb}\rightarrow\pi\eta^{\prime}}/\Gamma_{\pi_{1}%
^{hyb}\rightarrow\pi\eta}$ & $12.7\rule{0pt}{3.7ex}\rule[-2ex]{0pt}{0pt}%
$\\\hline
$\Gamma_{K_{1}^{hyb}\rightarrow K\eta}/\Gamma_{\pi_{1}^{hyb}\rightarrow\pi
\eta}$ & $0.69$ \rule{0pt}{3.7ex}\rule[-2ex]{0pt}{0pt}\\\hline
$\Gamma_{K_{1}^{hyb}\rightarrow K\eta^{\prime}}/\Gamma_{\pi_{1}^{hyb}%
\rightarrow\pi\eta}$ & $5.3\rule{0pt}{3.7ex}\rule[-2ex]{0pt}{0pt}$\\\hline
$\Gamma_{\eta_{1,N}^{hyb}\rightarrow\eta\eta^{\prime}}/\Gamma_{\pi_{1}%
^{hyb}\rightarrow\pi\eta}$ & $2.2$\\\hline
$\Gamma_{\eta_{1,S}^{hyb}\rightarrow\eta\eta^{\prime}}/\Gamma_{\pi_{1}%
^{hyb}\rightarrow\pi\eta}$ & $1.57$\\\hline
\end{tabular}
\caption{Ratios for the decay of vector hybrid mesons into two pseudoscalar
mesons (term proportional to $\beta_{A}^{hyb}$ in Eq. (\ref{hybriddecay})).}%
\label{tabIX}%
\end{table}

\section{Results}

\label{sec:results}

\subsection{Masses}

We compute the masses of vector and pseudovector hybrid mesons by using Eqs.
(\ref{mPi1}-\ref{mh1s}) in which $\pi_{1}$ is identified with $\pi_{1}(1600)$
(with mass $1660_{-11}^{+15}$ MeV) and the mass of $b_{1}^{hyb}$ is set to $2$
GeV. Moreover, we use $\delta_{S}^{hyb}\simeq\delta_{S}=0.151$ GeV$^{2}$.
Then, we obtain $h_{3}^{hyb}=-45.7$. The results for the other hybrid states
are reported in Table \ref{tabIII-IV}. We thus expect the other members of the
nonet of $\eta_{1,N}^{hyb}$, $\eta_{1,S}^{hyb},$ and $K_{1}^{hyb}$ to be also
well below $2$ GeV. Eventually, they can be also very broad, just as $\pi
_{1}(1600)$, rendering their experimental discovery quite challenging, but
-just as $\pi_{1}(1600)$- not impossible.

In particular, one may observe that our model predicts the state $\eta
_{1,N}^{hyb}$ to have the same mass of $\pi_{1}^{hyb}\equiv\pi_{1}(1600).$ The
nonstrange-strange mixing is expected to be small, since the hybrid mesons
under study are grouped into left- and right-handed current, and thus build a
so-called homochiral multiplet according to the classification of Ref.
\cite{Giacosa:2017pos}. We then do not expect a sizable shift of $\eta
_{1,N}^{hyb}$ and $\eta_{1,S}^{hyb}$ in Table \ref{tabIII-IV}.

In the lattice work of Ref. \cite{Dudek:2013yja} the isoscalar hybrids were
investigated together with the isovector state $\pi_{1}^{hyb}$. There is a
predominantly nonstrange hybrid meson which corresponds to $\eta_{1,N}^{hyb}$
state: its mass is about 100-150 MeV heavier than $\pi_{1}^{hyb}$ for $m_{\pi
}=391$ MeV. The value of this mass difference in the limit of the physical
pion mass is not yet settled. Our prediction concerning the similar mass about
$\eta_{1,N}^{hyb}$ and $\pi_{1}^{hyb}$ is upheld and should be verified in
future lattice evaluations. Admittedly, if the large mass difference presently
found in Ref. \cite{Dudek:2013yja} shall be confirmed in the future, it will
be hard to understand it in the framework of our model.

Moreover, there is a predominantly strange-antistrange hybrid meson about 300
MeV heavier than $\pi_{1}^{hyb}$, which corresponds to $\eta_{1,S}^{hyb}.$ It
is interesting to notice that the fields found on the lattice show a small but
nonzero nonstrange-strange mixing, that is not included in our model. The
small mixing is in agreement with the `homochiral' nature of the chiral
multiplet mentioned above. In the future, it will be also interesting to
include the effect of this mixing into the eLSM, which can generate a small
mass difference between $\eta_{1,N}^{hyb}$ and $\pi_{1}^{hyb}.$

In the lattice work of Ref. \cite{Dudek:2010wm} a kaonic hybrid, corresponding
to $K_{1}^{hyb}$, could be identified (the pion mass was about $400$ MeV). In
conclusions, there are robust lattice candidates for the whole nonet of hybrid
states \{$\pi_{1}^{hyb},$ $K_{1}^{hyb},\eta_{1,N}^{hyb},$ $\eta_{1,S}^{hyb}%
$\}. We regard this result as an important support of our model. New lattice
results about hybrid mesons with a pion mass approaching the physical value
would be very useful to confirm the overall picture and investigate eventual discrepancies.


\subsection{Decays}

The coupling constants \{$\lambda_{1}^{hyb},$ $\lambda_{2}^{hyb},$
$\alpha^{hyb}$, $\beta_{A}^{hyb}$\} entering the Lagrangian of Eq.
(\ref{hybriddecay}) are not known and cannot be determined as long as a clear
experimental information about (at least some of) the decay rates of hybrids
is missing. Nevertheless, we can build ratios of decays, since the values of
the coupling constant cancel out. Various decay ratios are reported in the
Tables \ref{tabIII-IV}-\ref{tabIX} and are independent on the parameters of
the model. In each Table, a reference decay has been chosen for building the
ratios. Any other desired ratio can be constructed by dividing entries in the tables.

In the present work, we restrict to large-$N_{c}$ dominant interaction terms
and we neglect decay terms that break flavor symmetry. Hence, the ratio that
we obtain are subject to these uncertainties and approximations. While in Ref.
\cite{Parganlija:2012fy} an agreement at the $10\%$ level was obtained, this
is definitely too optimistic in the present study of hybrids, which involves
heavier states and a poor experimental knowledge. Similarly to Ref.
\cite{Parganlija:2016yxq} that dealt with heavier states, we rather expect -as
an educated guess- an agreement of the order $20$-$30\%,$ but this value
should be seen only as a rough estimate, since a determination of the accuracy
is not possible at the present stage. At the same time, it must be remarked
that the aim of this work is not (and cannot be yet) a precise calculation of
hybrid decays, but rather the determination of some useful decay ratios that
can help toward the identification of possible hybrid candidates in the
(hopefully near) future. The results can tell us which decay channels are
favoured and which turns out to be suppressed according to the underlying
symmetries used in the model. In the future, when more precise data will be
available, one may study in more detail to which extent the underlying
symmetries (and their breaking patterns) are still applicable for these
unconventional mesonic states.

Next, we move to the presentation of each decay channel step by step. In (the
right part of) Table \ref{tabIII-IV} we report the decays of the $1^{-+}$ and
$1^{+-}$ hybrid states into a pseudovector and a pseudoscalar. The by far
dominant decay is $\pi_{1}^{hyb}\rightarrow b_{1}\pi$, that we use as our
reference decay. This decay mode should indeed one of the dominant decays of
the broad state $\pi_{1}(1600)$ that sizably contributes to the broad decay
width of this state. This is in agreement with the model results of Ref.
\cite{Close:1994hc} and the lattice results of Ref. \cite{McNeile:2006bz},

The second term of the Lagrangian (\ref{hybriddecay}) contains two- and
three-body decays. The dominant decay channel is $b_{1}^{hyb}\rightarrow\pi
\pi\eta.$ \ For what concerns the state $\pi_{1}^{hyb},$ one expects a quite
small decay $\pi_{1}^{hyb}\rightarrow K^{\ast}K\pi$, since the coupling is
proportional to the combination $\phi_{N}-\sqrt{2}\phi_{S},$ that vanishes in
the flavour limit $U_{V}(3)$ (where $\phi_{N}=\sqrt{2}\phi_{S}$). The decay
$\pi_{1}^{hyb}\rightarrow\bar{K}K$ vanishes even if an interaction Lagrangian
is present (see Appendix $C.2$ for details). This is in agreement with the
fact that $\pi_{1}^{hyb,0}\rightarrow\bar{K}K$ would violate $C$-parity, while
$\pi_{1}^{hyb,+}\rightarrow K^{+}K^{0}$ and $\pi_{1}^{hyb,-}\rightarrow
K^{-}K^{0}$ would violate $G$-parity. The by far largest decay of $\pi
_{1}^{hyb}$ for this term is the channel $\pi_{1}^{hyb}\rightarrow\pi\rho\eta$
(then, a $\pi\pi\pi\eta$ final state).

The third term of the Lagrangian (\ref{hybriddecay}) describes decays into
vector-pseudoscalar and axial-vector-pseudoscalar pairs. Two decays of
$\pi_{1}^{hyb}$ are expected to be sizable:
\begin{equation}
\pi_{1}^{hyb}\rightarrow\rho\pi\text{ and }\pi_{1}^{hyb}\rightarrow K^{\ast}K
\end{equation}
Other interesting and potentially large decays are $\eta_{1N}\rightarrow
K^{\ast}K,$ $\eta_{1N}\rightarrow K^{\ast}K,$ $b_{1}^{hyb}\rightarrow a_{1}%
\pi$ , see Table \ref{tabVII-VIII} for the full list.

The fourth and the last term describes the decays of the $1^{-+}$ hybrid nonet
states into two pseudoscalar states, one of which is either the $\eta$ or the
$\eta^{\prime}.$ The term explicitly breaks the axial symmetry $U_{A}(1)$
(although it preserves chiral symmetry, see the Appendix A), thus the flavor
blind state $\eta_{0}$ plays a crucial role. It is interesting to observe that
this term does not lead to two-body decays of the $1^{+-}$ hybrid states (see
Appendix \ref{appC4} for more details). In particular, the decays%
\begin{equation}
\pi_{1}^{hyb}\rightarrow\eta\pi\text{ and }\pi_{1}^{hyb}\rightarrow
\eta^{\prime}\pi
\end{equation}
are a consequence of this decay channel, with the decay channel $\pi_{1}%
^{hyb}\rightarrow\eta^{\prime}\pi$ being favoured (this is due to the fact
that $\eta^{\prime}$ is closer to the flavor singlet, while the meson $\eta$
is closer to the octet configuration): the ratio $\Gamma_{\pi_{1}%
^{hyb}\rightarrow\eta^{\prime}\pi}/\Gamma_{\pi_{1}^{hyb}\rightarrow\eta\pi}$
equals $12.7$ (a large ratio was also predicted in\ Ref. \cite{Bass:2001zs}).
At present, the decay modes $\pi_{1}(1600)\rightarrow\eta^{\prime}\pi$ and
$\pi_{1}(1400)\rightarrow\eta\pi$ have been observed. As already discussed, if
$\pi_{1}(1600)$ and $\pi_{1}(1400)$ corresponds to the same state
\cite{Rodas:2018owy} (see also Ref. \cite{Szczepaniak:2003vg}), then both
decay modes have been measured. The determination of the ratio in the future
would constitute an important test of our approach. The summary of the results
for this term are presented in Table \ref{tabIX}. Notice also that the decay
into the identical pair $\eta\eta$ does not take place, since the amplitude
for this process vanishes exactly when the direct and the crossed tree-level
Feynman diagrams are taken into account. This is expected because the two
identical $\eta$ mesons in $L=1$ configuration have positive parity, at odd
with the initial state which has negative parity; for more details on this
point, we refer to Appendix C.4.

As a last remark, we note that different decays listed in the tables can, by
further decay of some products, lead to the same final state. For instance, in
our calculations, the width $\Gamma_{b_{1}^{0,hyb}\rightarrow\pi^{+}a_{0}^{-}%
}$ is evaluated under the assumption that $a_{0}\equiv a_{0}(1450)$ is
stable.\ However, in reality $a_{0}(1450)$ is not stable and may decay further
into $\pi\eta,$ thus the following decay chain takes place: $b_{1}%
^{0,hyb}\rightarrow\pi^{+}a_{0}^{-}(1450)\rightarrow\pi^{+}\pi^{-}\eta.\ $On
the other hand, we also have the direct decay channel $b_{1}^{0,hyb}%
\rightarrow\pi^{+}\pi^{-}\eta.$ More in general, the following reactions
\begin{equation}
b_{1}^{0,hyb}\rightarrow\pi^{+}\pi^{-}\eta\text{ , }b_{1}^{0,hyb}%
\rightarrow\pi^{+}a_{0}^{-}\rightarrow\pi^{+}\pi^{-}\eta\text{ },\text{ }%
b_{1}^{0,hyb}\rightarrow\pi^{-}a_{0}^{+}\rightarrow\pi^{-}\pi^{+}\eta\text{
,...}%
\end{equation}
end up in the same final state, where dots refer to other possible decay
chains. In principle, one should first perform the sum of the related
amplitudes and the square, obtaining:\textbf{ }%
\begin{equation}
\Gamma_{b_{1}^{0,hyb}\rightarrow\pi^{+}\pi^{-}\eta}^{\text{total}}%
=\Gamma_{b_{1}^{0,hyb}\rightarrow\pi^{+}\pi^{-}\eta}^{\text{direct}}%
+\Gamma_{b_{1}^{0,hyb}\rightarrow\pi^{+}a_{0}^{-}\rightarrow\pi^{+}\pi^{-}%
\eta}+\Gamma_{b_{1}^{0,hyb}\rightarrow\pi^{-}a_{0}^{+}\rightarrow\pi^{-}%
\pi^{+}\eta}+...+\text{ \textquotedblleft mixed terms\textquotedblright,}%
\end{equation}
where mixed terms refer to interference effects between the amplitudes.
Fortunately, such interference effects are typically small, since the overlap
of different channels is suppressed, as the explicit calculation presented in
Ref. \cite{Eshraim:2012rb} shows. More in details, one can study the
distributions in the Dalitz plot $\pi^{+}\pi^{-}\eta$ in order to distinguish
the contributions. Summarizing, in our work we do not consider these effects,
since they are not expected to sizably change the result and they go beyond
the accuracy of our tree-level approach. Yet, they should be included once
concrete candidates and accurate experimental results will be available.

\section{Conclusions and Outlook}

\label{sec:conclusions}

In this work we have studied masses and decays of the lightest hybrid nonet
with $J^{PC}=1^{-+}$ and of its chiral partner nonet with $J^{PC}=1^{+-}.$ To
this end, we have embedded the hybrid state into a chiral multiplet and
coupled it to the chiral model called eLSM. Upon assigning the resonance
$\pi_{1}(1600)$ to the isovector member of the lightest hybrid nonet , we have
made predictions for some masses of hybrid states and for branching ratios of
$\pi_{1}(1600)$ and the members of this multiplet as well as the for their
chiral partners. The main results are reported in Tables \ref{tabIII-IV}%
-\ref{tabIX}.

For what concerns the masses, there are three hybrid states with
$J^{PC}=1^{-+}$ , denoted as $K_{1}$ as well as $\eta_{1,N}$ and $\eta_{1,S}.$
Their discovery is then possible, provided that these states are not too wide.
For what concerns decays, we have introduced four chirally invariant effective
interaction terms describing the masses and the two- and three-body decays of hybrids.

The interaction Lagrangian describing the hybrid-meson decays into other
mesons is presented in Eq. (\ref{hybriddecay}). The first and the second terms
in the interaction Lagrangian fulfill the chiral and dilatation symmetries and
for this reason are expected to deliver the dominant contributions to the
decays of the hybrid states. In particular, the first term of our approach
describes decays of the $J^{PC}=1^{-+}$ state into pseudovector ($J^{PC}%
=1^{+-}$) and pseudoscalar states, such as $\pi_{1}(1600)\rightarrow
b_{1}(1230)\pi\rightarrow\omega\pi\pi$. Hence, the final state $\omega\pi\pi$
represents a promising channel for the confirmation of this hybrid candidate.
Analogous decays of the other exotic hybrids have been obtained as a
prediction. In addition, the decays of crypto-exotic hybrids into the scalar
and orbitally excited vector mesons could be evaluated. According to the
second term, $\pi_{1}(1600)$ decays into $KK\pi$, $\rho\pi\eta,$ and $KK$, but
only the decay into $\rho\pi\eta$ is expected to be sizable. The third term of
the interaction Lagrangian breaks dilatation invariance and generates also
three-body and two-body decays. The latter are important, since they contain
the process $\pi_{1}\rightarrow\rho\pi\rightarrow\pi\pi\pi$, thanks to which
the $\pi_{1}(1600)$ was seen at COMPASS. Decays of other member of the
multiplet and their chiral partners are presented as predictions. Finally, the
decays $\pi_{1}\rightarrow\eta\pi$ and $\pi_{1}\rightarrow\eta^{\prime}\pi$
emerge from the fourth Lagrangian term which breaks axial and dilatation
symmetries (but still fulfills chiral symmetry). These decay modes, even if
subleading, are seen in experiment due to the very clean nature of their decay
products. It is quite remarkable that, within our setup, the only way to
obtain such decays goes through the axial anomaly.\ A breaking of flavor
symmetry in the first three decay terms does not lead to decays into $\eta\pi$
and $\eta^{\prime}\pi$.

Summarizing, for the resonance $\pi_{1}(1600)$ we expect the following
decays:
\begin{equation}
\pi_{1}(1600)\rightarrow\pi b_{1}\text{, }\pi_{1}(1600)\rightarrow\rho\pi
\eta\text{, }\pi_{1}(1600)\rightarrow\rho\pi\text{ , }\pi_{1}(1600)\rightarrow
K^{\ast}(892)K\text{ , }\pi_{1}(1600)\rightarrow\eta^{\prime}\pi,\text{ }%
\pi_{1}(1600)\rightarrow\eta\pi\text{.}%
\end{equation}
It is hard at the present stage to determine which one of them is the largest.
We expect the $\pi b_{1}$ mode to be quite large, in agreement with the
lattice study of Ref. \cite{McNeile:2006bz}. At the same time, the latter two
decays are expected to be small, since they break explicitly the axial anomaly.

At present, the results of these paper are at the tree-level. As a possible
outlook, one can calculate the spectral function of $\pi_{1}(1600).$ One may
start with the dominant terms discussed in this work and calculate loops,
following the techniques described in\ Ref.
\cite{Tornqvist:1995kr,Tornqvist:1995ay,Boglione:1997aw,Boglione:2002vv,Wolkanowski:2015lsa,Wolkanowski:2015jtc}%
. This can be quite important, as shown in the recent work of Ref.
\cite{Rodas:2018owy}. Another possibility is to study other hybrid nonets
(such as for instance tensor hybrids) by repeating the steps presented in this work.

Summarizing, the confirmation of $\pi_{1}(1600)$ as a genuine hybrid state as
well as the discovery of the other members of the nonet and its chiral
partners would represent a step forward in our understanding of QCD, for which
both theoretical and experimental efforts are worth to be spent.

\section*{Acknowledgments}

The authors are grateful for discussions with F. Maas, D. H. Rischke. W.I.E.
and C.S.F. acknowledge support from the BMBF under contracts No. 05H15RGKBA as
well as from HIC for FAIR. C.S.F. acknowledges support from the BMBF under
contract No. 05P18RGFP1. F.G. acknowledges support from the Polish National
Science Centre (NCN) through the OPUS projects no. 2019/33/B/ST2/00613 and no. 2018/29/B/ST2/02576.

\appendix

\section{Details of the eLSM}

\label{appA}

The Lagrangian of the eLSM for (pseudo)scalar and (axial-)vector states,
constructed upon requiring chiral symmetry ($U(3)_{R}\times U(3)_{L}$),
dilatation invariance, as well as under charge conjugation $C$ and parity $P$
symmetries, reads:
\begin{align}
\mathcal{L}_{eLSM}  &  =%
\mathcal{L}%
_{dil}+\mathrm{Tr}[(D_{\mu}\Phi)^{\dagger}(D^{\mu}\Phi)]-m_{0}^{2}\left(
\frac{G}{G_{0}}\right)  ^{2}\mathrm{Tr}(\Phi^{\dagger}\Phi)-\lambda
_{1}[\mathrm{Tr}(\Phi^{\dagger}\Phi)]^{2}-\lambda_{2}\mathrm{Tr}(\Phi
^{\dagger}\Phi)^{2}\nonumber\\
&  -\frac{1}{4}\mathrm{Tr}[(L^{\mu\nu})^{2}+(R^{\mu\nu})^{2}]+\mathrm{Tr}%
\left[  \left(  \frac{m_{1}^{2}}{2}\left(  \frac{G}{G_{0}}\right)  ^{2}%
+\Delta\right)  (L_{\mu}^{2}+R_{\mu}^{2})\right]  +\mathrm{Tr}[H(\Phi
+\Phi^{\dagger})]\nonumber\\
&  +c_{1}(\mathrm{det}\Phi-\mathrm{det}\Phi^{\dagger})^{2}+i\frac{g_{2}}%
{2}\{\mathrm{Tr}(L_{\mu\nu}[L^{\mu},L^{\nu}])+\mathrm{Tr}(R_{\mu\nu}[R^{\mu
},R^{\nu}])\}\nonumber\\
&  +\frac{h_{1}}{2}\mathrm{Tr}(\Phi^{\dagger}\Phi)\mathrm{Tr}\left(  L_{\mu
}^{2}+R_{\mu}^{2}\right)  +h_{2}\mathrm{Tr}[\left\vert L_{\mu}\Phi\right\vert
^{2}+\left\vert \Phi R_{\mu}\right\vert ^{2}]\nonumber\\
&  +2h_{3}\mathrm{Tr}(L_{\mu}\Phi R^{\mu}\Phi^{\dagger})+\mathcal{L}%
_{eLSM}^{\tilde{\Phi}}...\text{ ,} \label{fulllag}%
\end{align}
where $D^{\mu}\Phi=\partial^{\mu}\Phi-ig_{1}(L^{\mu}\Phi-\Phi R^{\mu})$ and
the dilaton (i.e. the scalar glueball) Lagrangian is
\begin{equation}%
\mathcal{L}%
_{dil}=\frac{1}{2}(\partial_{\mu}G)^{2}-\frac{1}{4}\frac{m_{G}^{2}}%
{\Lambda^{2}}\left(  G^{4}\ln\left\vert \frac{G}{\Lambda}\right\vert
-\frac{G^{4}}{4}\right)  \text{ ,}%
\end{equation}
see Refs.\ \cite{Parganlija:2012fy,Janowski:2014ppa} for details.\textbf{ }We
recall that the two diagonal matrices $H$ and $\Delta$ parametrize the
explicit breaking of chiral symmetry due to nonzero quark masses. Moreover,
the term proportional to $c_{1}$ describes the axial anomaly. Finally,
$\mathcal{L}_{eLSM}^{\tilde{\Phi}}$ contains the kinetic as well as
interaction terms for the chiral multiplet $\tilde{\Phi}^{\mu}=V_{E}^{\mu
}-iB^{\mu},$ whose detailed form was not yet unexplored (but is not relevant
for us).

Thank to $%
\mathcal{L}%
_{dil},$ one can describe the breaking of dilatation invariance. In the chiral
limit ($\Delta=H=0$) and neglecting terms linked to the axial anomaly
($c_{1}=0$ and in the hybrid sector $\alpha^{hyb}=\beta_{A}^{hyb}=0)$, the
parameter $\Lambda$ is the only dimensionful parameter of Eq. (\ref{fulllag}),
Namely, all other quantities are described by dimensionless parameters, e.g.
$h_{k},$ $h_{k}^{hyb}$ with $k=1,2,3.$ In this way only a finite number of
terms is possible. The breaking of dilatation invariance due to quark masses,
parametrized by $H$ and $G,$ is rather small, but terms describing the chiral
anomaly may be non-negligible.

The field $G$ develops a nonzero vacuum's expectation value $G_{0}$ (note,
$G_{0}=\Lambda$ in the limit in which the glueball decouples from
(pseudo)scalar fields, i.e. $m_{0}=0),$ hence a shift is needed:
\begin{equation}
G\rightarrow G_{0}+G\text{ .}%
\end{equation}

Next, for $m_{0}^{2}<0$ (realized in Nature), spontaneous breaking of chiral
symmetry takes place.\ As a consequence, one has to perform the shift of the
scalar-isoscalar quark-antiquark fields by their vacuum expectation values
$\phi_{N}$ and $\phi_{S}$:
\begin{equation}
\sigma_{N}\rightarrow\sigma_{N}+\phi_{N}\text{ and }\sigma_{S}\rightarrow
\sigma_{S}+\phi_{S}\text{ .} \label{shift}%
\end{equation}
In matrix form:%
\begin{equation}
S\rightarrow\Phi_{0}+S\text{ with }\Phi_{0}=\frac{1}{\sqrt{2}}\left(
\begin{array}
[c]{ccc}%
\frac{\phi_{N}}{\sqrt{2}} & 0 & 0\\
0 & \frac{\phi_{N}}{\sqrt{2}} & 0\\
0 & 0 & \phi_{S}%
\end{array}
\right)  \text{ .} \label{cond}%
\end{equation}
Note, one can rewrite $\Phi_{0}$ as
\begin{equation}
\Phi_{0}=\frac{\phi_{N}}{2}1_{3}+\left(  \frac{\phi_{S}}{\sqrt{2}}-\frac
{\phi_{N}}{2}\right)  diag\{0,0,1\}\text{, } \label{apprphi}%
\end{equation}
where the first term is dominant and the second is a flavour breaking
correction since $\phi_{N}\simeq\sqrt{2}\phi_{S}.$ In addition, one has also
to `shift' the axial-vector fields%
\begin{align}
\vec{a}_{1}^{\mu}  &  \rightarrow\vec{a}_{1}^{\mu}+Z_{\pi}w_{\pi}\partial
^{\mu}\vec{\pi}\text{ , }K_{1,A}^{+,\mu}\rightarrow K_{1,A}^{+,\mu}+Z_{K}%
w_{k}\partial^{\mu}K\text{, ...}\nonumber\\
f_{1,N}^{\mu}  &  \rightarrow f_{1,N}^{\mu}+Z_{\eta_{N}}w_{\eta_{N}}%
\partial^{\mu}\eta_{N}\text{ , }f_{1,S}^{\mu}\rightarrow f_{1,S}^{\mu}%
+Z_{\eta_{S}}w_{\eta_{S}}\partial^{\mu}\eta_{S}\text{ ,}%
\end{align}
and to consider the wave-function renormalization of the pseudoscalar fields:%
\begin{equation}
\vec{\pi}\rightarrow Z_{\pi}\vec{\pi}\text{ , }K^{+}\rightarrow Z_{K}%
K^{+}\text{, ...} \label{shifta}%
\end{equation}%
\begin{equation}
\eta_{N}\rightarrow Z_{\eta_{N}}\eta_{N}\;,\eta_{S}\rightarrow Z_{\eta_{S}%
}\eta_{S}\text{ .} \label{shiftb}%
\end{equation}
The constants entering into the previous expressions are:%
\begin{equation}
Z_{\pi}=Z_{\eta_{N}}=\frac{m_{a_{1}}}{\sqrt{m_{a_{1}}^{2}-g_{1}^{2}\phi
_{N}^{2}}}\text{ , }Z_{K}=\frac{2m_{K_{1,A}}}{\sqrt{4m_{K_{1,A}}^{2}-g_{1}%
^{2}(\phi_{N}+\sqrt{2}\phi_{S})^{2}}}\text{ , }Z_{\eta_{S}}=\frac{m_{f_{1S}}%
}{\sqrt{m_{f_{1S}}^{2}-2g_{1}^{2}\phi_{S}^{2}}}\;\text{,} \label{zpi}%
\end{equation}
and:%
\begin{equation}
w_{\pi}=w_{\eta_{N}}=\frac{g_{1}\phi_{N}}{m_{a_{1}}^{2}}\;\text{,}\quad
w_{K}=\frac{g_{1}(\phi_{N}+\sqrt{2}\phi_{S})}{2m_{K_{1},A}^{2}}\text{ ,
}w_{\eta_{S}}=\frac{\sqrt{2}g_{1}\phi_{S}}{m_{f_{1S}}^{2}}\;\;\text{.}
\label{wf1}%
\end{equation}
The numerical values of the renormalization constants are $Z_{\pi}=1.709$,
$Z_{K}=1.604,$ $Z_{\eta_{S}}=1.539$ \cite{Parganlija:2012fy}, while those of
the $w$-parameters are: $w_{\pi}=0.683$ GeV$^{-1},$ $w_{K}=0.611$ GeV$^{-1}$ ,
$w_{\eta_{S}}=0.554$ GeV$^{-1}$. Moreover, the condensates $\phi_{N}$ and
$\phi_{S}$ read%
\begin{equation}
\phi_{N}=Z_{\pi}f_{\pi}=0.158\text{ GeV, }\phi_{S}=\frac{2Z_{K}f_{K}-\phi_{N}%
}{\sqrt{2}}=0.138\text{ GeV}\;\text{,}%
\end{equation}
where the standard values $f_{\pi}=0.0922$ GeV and $f_{K}=0.110$ GeV have been
used \cite{Tanabashi:2018oca}. The previous expressions can be summarized by
the matrix replacements
\begin{equation}
P\rightarrow\mathcal{P}=\frac{1}{\sqrt{2}}\left(
\begin{array}
[c]{ccc}%
\frac{Z_{\pi}}{\sqrt{2}}(\eta_{N}+\pi^{0}) & Z_{\pi}\pi^{+} & Z_{K}K^{+}\\
Z_{\pi}\pi^{-} & \frac{Z_{\pi}}{\sqrt{2}}(\eta_{N}-\pi^{0}) & Z_{K}K^{0}\\
Z_{K}K^{-} & Z_{K}\bar{K}^{0} & Z_{\eta_{S}}\eta_{S}%
\end{array}
\right)  \text{ ,}%
\end{equation}
and%
\begin{equation}
A^{\mu}\rightarrow\mathcal{A}^{\mu}=\frac{1}{\sqrt{2}}\left(
\begin{array}
[c]{ccc}%
\frac{f_{1N}+a_{1}^{0}}{\sqrt{2}} & a_{1}^{+} & K_{1,A}^{+}\\
a_{1}^{-} & \frac{f_{1N}-a_{1}^{0}}{\sqrt{2}} & K_{1,A}^{0}\\
K_{1,A}^{-} & \bar{K}_{1,A}^{0} & f_{1S}%
\end{array}
\right)  ^{\mu}+\frac{\partial^{\mu}}{\sqrt{2}}\left(
\begin{array}
[c]{ccc}%
\frac{Z_{\pi}w_{\pi}}{\sqrt{2}}(\eta_{N}+\pi^{0}) & Z_{\pi}w_{\pi}\pi^{+} &
Z_{K}w_{K}K^{+}\\
Z_{\pi}w_{\pi}\pi^{-} & \frac{Z_{\pi}w_{\pi}}{\sqrt{2}}(\eta_{N}-\pi^{0}) &
Z_{K}w_{K}K^{0}\\
Z_{K}w_{K}K^{-} & Z_{K}w_{K}\bar{K}^{0} & Z_{\eta_{S}}w_{\eta_{S}}\eta_{S}%
\end{array}
\right)  \text{ .} \label{shifttot}%
\end{equation}
Note, in the $U_{V}(3)$ limit (in which all three bare quark masses are
equals) some simplifications take place (useful for cross-check of the
results): $\Phi_{N}=\sqrt{2}\Phi_{S}$, $Z=Z_{\pi}=Z_{K}=Z_{\eta_{S}}$, and
$w=w_{\pi}=w_{K}=w_{\eta_{S}}$, out of which $P\rightarrow P=ZP$ and $A^{\mu
}\rightarrow A^{\mu}=A^{\mu}+Zw\partial^{\mu}P$.

The chiral anomaly is described by the term $c_{1}(det\Phi-det\Phi^{\dagger
})^{2}$ in Eq. (\ref{fulllag}). This term is not invariant under
$U_{3}(R)\times U_{3}(R)$, according to which $\Phi\rightarrow U_{L}\Phi
U_{R}^{\dagger}$ (See Table 2). In fact, it is only invariant under
$SU_{3}(R)\times SU_{3}(R)$, for which $\det U_{L}=\det U_{R}=1.$ If however
we consider the $U_{A}(1)$ transformation $\Phi\rightarrow e^{i\alpha}\Phi$
(obtained for $U_{L}=$ $1_{3}e^{i\alpha/2}$ and $U_{R}=U_{L}^{\dagger}),$ the
determinant implies that this symmetry is broken: in this way the $U_{A}(1)$
anomaly is taken into account. As a consequence, this term generates a term
proportional to $\eta_{0}^{2},$ which shifts up the mass of the flavor singlet
configuration, in a way analogous to the one originally described in Ref.
\cite{tHooft:1986ooh}. Thus, the masses of $\eta$ and $\eta^{\prime}$ can be
correctly described \cite{Parganlija:2012fy}. For more details and other
anomalous terms, see Ref. \cite{Giacosa:2017pos,Fariborz:2005gm}.

The eLSM has been enlarged to four flavors
in\ Refs.~\cite{Eshraim:2014eka,Eshraim:2018iea}. Interestingly, charmed meson
masses and large-$N_{c}$ dominant decays can be described relatively well
(even if one is far from the natural domain of chiral symmetry). In the end,
we also recall that the pseudoscalar glueball can be coupled to the eLSM via
the chiral Lagrangian $%
\mathcal{L}%
_{\tilde{G}}=ic_{\tilde{G}\Phi}\tilde{G}\left(  \text{\textrm{det}}%
\Phi-\text{\textrm{det}}\Phi^{\dag}\right)  $, which reflects the axial
anomaly in the pseudoscalar-isoscalar sector, see details and results in
Refs.\ \cite{Eshraim:2012jv,Eshraim:2016mds}. In a recent extension, the very
same Lagrangian is used to study the decay of an hypothetical excited
pseudoscalar glueball \cite{Eshraim:2019sgr}.

\section{Absence of shift for vector hybrid states}

\label{appB}

There is no allowed term which mixes the hybrid nonets with (pseudo)scalar
mesons. Namely, one may start from the general chirally invariant Lagrangian
term involving hybrid fields as well as $\Phi$ and $\partial^{\mu}\Phi$
\begin{equation}
\mathcal{L}_{\text{test}}=\alpha\mathrm{Tr}[\left(  \partial^{\mu}\Phi\right)
\Phi^{\dagger}L_{\mu}^{hyb}]+\beta\mathrm{Tr}[\left(  \partial^{\mu}%
\Phi\right)  R_{\mu}^{hyb}\Phi^{\dag}] \label{hphiphi}%
\end{equation}
Note, other terms can be always recasted in a combination of the previous one.
For instance%
\begin{align}
\mathrm{Tr}[\left(  \partial^{\mu}\Phi^{\dag}\right)  L_{\mu}^{hyb}\Phi]  &
=\mathrm{Tr}\left[  \partial^{\mu}\left(  \Phi^{\dag}L_{\mu}^{hyb}\Phi\right)
\right]  -\mathrm{Tr}\left[  \Phi^{\dag}\left(  \partial^{\mu}L_{\mu}%
^{hyb}\right)  \Phi\right]  -\mathrm{Tr}\left[  \Phi^{\dag}L_{\mu}%
^{hyb}\left(  \partial^{\mu}\Phi\right)  \right] \nonumber\\
&  \equiv-\mathrm{Tr}[\Phi^{\dag}L_{\mu}^{hyb}\left(  \partial^{\mu}%
\Phi\right)  ]=-\mathrm{Tr}[\left(  \partial^{\mu}\Phi\right)  \Phi^{\dagger
}L_{\mu}^{hyb}]
\end{align}
where a full derivative has been neglected and $\partial^{\mu}R_{\mu}%
^{hyb}=\partial^{\mu}L_{\mu}^{hyb}=0$ (since they are divergenceless vector fields).

Under parity transformation:%
\begin{align}
&  \mathcal{L}_{\text{test}}\overset{P}{\rightarrow}\alpha\mathrm{Tr}[\left(
\partial^{\mu}\Phi^{\dagger}\right)  \Phi R_{\mu}^{hyb}]+\beta\mathrm{Tr}%
[\left(  \partial^{\mu}\Phi^{\dag}\right)  L_{\mu}^{hyb}\Phi]=-\alpha
\mathrm{Tr}[\Phi^{\dagger}\left(  \partial^{\mu}\Phi\right)  R_{\mu}%
^{hyb}]-\beta\mathrm{Tr}[\Phi^{\dag}L_{\mu}^{hyb}\left(  \partial^{\mu}%
\Phi\right)  ]\nonumber\\
&  =-\beta\mathrm{Tr}[\left(  \partial^{\mu}\Phi\right)  \Phi^{\dagger}L_{\mu
}^{hyb}]-\alpha\mathrm{Tr}[\left(  \partial^{\mu}\Phi\right)  R_{\mu}%
^{hyb}\Phi^{\dag}]
\end{align}
where again similar manipulations have been applied. Therefore, if we impose
parity invariance, the condition $\beta=-\alpha$ follows.

Next, we consider \ $C$-parity, according to which Eq. (\ref{hphiphi})
transforms into%
\begin{align}
&  \mathcal{L}_{\text{test}}\overset{C}{\rightarrow}\alpha\mathrm{Tr}[\left(
\partial^{\mu}\Phi^{t}\right)  \left(  \Phi^{\dagger}\right)  ^{t}(R_{\mu
}^{hyb})^{t}]+\beta\mathrm{Tr}[\left(  \partial^{\mu}\Phi^{t}\right)  (L_{\mu
}^{hyb})^{t}\left(  \Phi^{\dag}\right)  ^{t}]=\alpha\mathrm{Tr}\left[  \left(
R_{\mu}^{hyb}\Phi^{\dagger}\left(  \partial^{\mu}\Phi\right)  \right)
^{t}\right]  +\beta\mathrm{Tr}\left[  \left(  \Phi^{\dag}L_{\mu}^{hyb}\left(
\partial^{\mu}\Phi\right)  \right)  ^{t}\right] \nonumber\\
&  =\beta\mathrm{Tr}[\left(  \partial^{\mu}\Phi\right)  \Phi^{\dagger}L_{\mu
}^{hyb}]+\alpha\mathrm{Tr}[\left(  \partial^{\mu}\Phi\right)  R_{\mu}%
^{hyb}\Phi^{\dag}]\text{ ,}%
\end{align}
out of which $\beta=\alpha$ assures invariance under $C.$

It is then clear that the only solution is
\begin{equation}
\alpha=\beta=0\text{ }\rightarrow\mathcal{L}_{\text{test}}=0\text{ ,}%
\end{equation}
i.e. the simultaneous requirement of invariance under $P$ and \ $C$ cannot be
fulfilled. In particular, it is the different transformation of hybrids under
$C$-parity that forbids this interaction. The only interaction involving one
hybrid field and two (pseudo)scalar ones does not contain derivatives and is
the one of Eq. (\ref{hybriddecay}).

The implications are important: there is no mixing such as the $a_{1}\pi$ one
discussed above. The fields entering Eq. (\ref{hybridmatrices}) are already
the physical ones.

\section{Decay rates for hybrid mesons}

\label{appC}

We present the explicit expressions for the two- and three-body decay rates
for hybrid mesons.

\subsection{First term of the Lagrangian of Eq. (\ref{hybriddecay})}

The first term of the effective Lagrangian (\ref{hybriddecay})
\[
\mathcal{L}_{eLSM,1}^{\text{ hybrid-linear}}=i\lambda_{1}^{hyb}G\mathrm{Tr}%
\left[  L_{\mu}^{hyb}(\tilde{\Phi}^{\mu}\Phi^{\dagger}-\Phi\tilde{\Phi}%
^{\dag\mu})+R_{\mu}^{hyb}(\tilde{\Phi}^{\mu\dag}\Phi-\Phi^{\dag}\tilde{\Phi
}^{\mu})\,\right]  \text{ }%
\]
describes the interaction of hybrid mesons with pseudovector and excited
vector mesons and (pseudo)scalar mesons.

Let us first verify the invariance under $P$ and $C.$ Under parity the first
term transforms as:%
\begin{equation}
\mathrm{Tr}(L_{\mu}^{hyb}\tilde{\Phi}^{\mu}\Phi^{\dagger})\overset
{P}{\rightarrow}\mathrm{Tr}(R^{hyb,\mu}\tilde{\Phi}_{\mu}^{\dag}\Phi)\text{ ,}%
\end{equation}
which equals the third term; similarly, the second converts into the fourth,
hence invariance under $P$ is guaranteed.

Next, under $C$ the first term transforms as:%
\begin{equation}
\mathrm{Tr}(L_{\mu}^{hyb}\tilde{\Phi}^{\mu}\Phi^{\dagger})\overset
{C}{\rightarrow}-\mathrm{Tr}(R_{\mu}^{hyb,t}\tilde{\Phi}^{\mu,t}\Phi^{\dagger
t})=-\mathrm{Tr}(\Phi^{\dagger}\tilde{\Phi}^{\mu}R_{\mu}^{hyb})=-\mathrm{Tr}%
(R_{\mu}^{hyb}\Phi^{\dagger}\tilde{\Phi}^{\mu})\text{ ,}%
\end{equation}
hence the first term converts into the fourth and the second into the
third.\ Invariance under is also fulfilled.

As a last check, we show that the Lagrangian is Hermitian. For the first term
(including the $i$ in front), one has:%

\begin{align}
\left\{  i\mathrm{Tr}\left[  L_{\mu}^{hyb}(\tilde{\Phi}^{\mu}\Phi^{\dagger
}-\Phi\tilde{\Phi}^{\dag\mu})\right]  \right\}  ^{\dag}  &  =-i\mathrm{Tr}%
\left[  (\tilde{\Phi}^{\mu}\Phi^{\dagger}-\Phi\tilde{\Phi}^{\dag\mu})^{\dag
}L_{\mu}^{hyb,\dag}\right]  =\mathrm{Tr}\left[  L_{\mu}^{hyb,\dag}(\Phi
\tilde{\Phi}^{\mu,\dag}-\tilde{\Phi}^{\mu}\Phi^{\dag})\right]  =\\
&  =-i\mathrm{Tr}\left[  L_{\mu}^{hyb}(\Phi\tilde{\Phi}^{\mu,\dag}-\tilde
{\Phi}^{\mu}\Phi^{\dag})\right]  =i\mathrm{Tr}\left[  L_{\mu}^{hyb}%
(\tilde{\Phi}^{\mu}\Phi^{\dagger}-\Phi\tilde{\Phi}^{\dag\mu})\right]  .
\end{align}
A similar expressions holds for the second term.

In terms of the physical nonets with defined $J^{PC}$, the Lagrangian can be
rewritten as:%

\begin{align}
\mathcal{L}_{eLSM,1}^{\text{ hybrid-linear}}  &  =i2\lambda_{1}^{hyb}G\left\{
\mathrm{Tr}\left[  \Pi_{\mu}^{hyb}\,[P,B^{\mu}]\right]  +\mathrm{Tr}\left[
\Pi_{\mu}^{hyb}\,[V_{E}^{\mu},S]\right]  \right\}  +\nonumber\\
&  2\lambda_{1}^{hyb}G\left\{  \mathrm{Tr}\left[  B_{\mu}^{hyb}\,\left\{
P,V_{E}^{\mu}\right\}  \right]  +\mathrm{Tr}\left[  B_{\mu}^{hyb}\,\left\{
B^{\mu},S\right\}  \right]  \right\}  .
\end{align}
This expression shows that following decays for the hybrid nonet $\Pi_{\mu
}^{hyb}$ are possible
\begin{equation}
\Pi_{\mu}^{hyb}\,\rightarrow PB^{\mu}\text{ and }\Pi_{\mu}^{hyb}\,\rightarrow
SV_{E}^{\mu}\text{ .}%
\end{equation}
However, the second is not relevant for our purposes because the corresponding
decay channels are kinematically forbidden.

For the hybrid nonet $B_{\mu}^{hyb}$ we get%
\begin{equation}
B_{\mu}^{hyb}\,\rightarrow PV_{E}^{\mu}\text{ and }\Pi_{\mu}^{hyb}%
\,\rightarrow B^{\mu}S\text{ ,}%
\end{equation}
where, as above, the second term leads to kinematically forbidden decays.

Note, for completeness we check also the invariance $i2\lambda_{1}%
^{hyb}G\mathrm{Tr}\left[  \Pi_{\mu}^{hyb}\,[P,B^{\mu}]\right]  $ under $P$ and
$C$ and $\dagger$:%
\begin{align}
\mathrm{Tr}\left[  \Pi_{\mu}^{hyb}\,[P,B^{\mu}]\right]  \overset
{P}{\rightarrow}\mathrm{Tr}\left[  \Pi^{hyb,\mu}\,[-P,-B_{\mu}]\right]   &
=\mathrm{Tr}\left[  \Pi_{\mu}^{hyb}\,[P,B^{\mu}]\right]  \text{ ;}\\
\mathrm{Tr}\left[  \Pi_{\mu}^{hyb}\,[P,B^{\mu}]\right]  \overset
{C}{\rightarrow}\mathrm{Tr}\left[  \Pi_{\mu}^{hyb,t}\,[P^{t},-B^{\mu
,t}]\right]   &  =-\mathrm{Tr}\left[  \Pi_{\mu}^{hyb,t}\,\,[P,B^{\mu}%
]^{t}\right]  =\mathrm{Tr}\left[  \Pi_{\mu}^{hyb}\,[P,B^{\mu}]\right]  \text{
;}\\
\left\{  i\mathrm{Tr}\left[  \Pi_{\mu}^{hyb}\,[P,B^{\mu}]\right]  \right\}
^{\dag}  &  =-i\mathrm{Tr}\left[  \,[P,B^{\mu}]^{\dag}\Pi_{\mu}^{hyb}\right]
=i\mathrm{Tr}\left[  \Pi_{\mu}^{hyb}\,[P,B^{\mu}]\right]  \text{ .}%
\end{align}
Similar check for the dominant decay term of $B_{\mu}^{hyb}$ are also reported:%

\begin{align}
\mathrm{Tr}\left[  B_{\mu}^{hyb}\,\left\{  P,V_{E}^{\mu}\right\}  \right]
\overset{P}{\rightarrow}\mathrm{Tr}\left[  -B^{hyb,\mu}\,\left\{  -P,V_{E,\mu
}\right\}  \right]   &  =\mathrm{Tr}\left[  B_{\mu}^{hyb}\,\left\{
P,V_{E}^{\mu}\right\}  \right]  \text{ ;}\\
\mathrm{Tr}\left[  B_{\mu}^{hyb}\,\left\{  P,V_{E}^{\mu}\right\}  \right]
\overset{C}{\rightarrow}\mathrm{Tr}\left[  -B_{\mu}^{hyb,t}\,\left\{
P^{,t},-V_{E}^{\mu,t}\right\}  \right]   &  =\mathrm{Tr}\left[  B_{\mu}%
^{hyb}\,\left\{  P,V_{E}^{\mu}\right\}  \right]  \text{ ;}\\
\left\{  \mathrm{Tr}\left[  B_{\mu}^{hyb}\,\left\{  P,V_{E}^{\mu}\right\}
\right]  \right\}  ^{\dag}  &  =\mathrm{Tr}\left[  \,\left\{  P,V_{E}^{\mu
}\right\}  ^{\dag}B_{\mu}^{hyb}\right]  =\mathrm{Tr}\left[  B_{\mu}%
^{hyb}\,\left\{  P,V_{E}^{\mu}\right\}  \right]  \text{ .}%
\end{align}

After performing the field transformations in Eq.(\ref{shift}), Eq.
(\ref{shifta}) and Eq.(\ref{shiftb}), it is calculate the corresponding terms
describing the decays. For instance, for the case of the state $\pi_{1}%
^{hyb}\pi b_{1}$ interaction, the following explicit Lagrangian term:%

\begin{equation}
\mathcal{L}_{eLSM-\lambda_{1}^{hyb}}^{\text{ hybrid-linear}}=\lambda_{1}%
^{hyb}G_{0}Z_{\pi}(\pi^{-}b_{1}^{0\mu}+\pi^{0}b_{1}^{-\mu})\pi_{1\mu}%
^{+hyb}+...\nonumber
\end{equation}
Then, the average modulus squared decay amplitude is given by%

\begin{equation}
|\mathcal{M}_{\pi_{1\mu}^{+hyb}\rightarrow b_{1}^{0}\pi^{+}}|^{2}=\frac{1}%
{3}G_{0}^{2}{\lambda_{1}^{hyb}}^{2}Z_{\pi}^{2}\left[  2+\frac{(m_{\pi
_{1}^{hyb}}^{2}+m_{b_{1}}^{2}-m_{\pi}^{2})^{2}}{4m_{A}^{2}m_{B_{1}}^{2}%
}\right]  \text{ ,}%
\end{equation}
hence the decay width reads:%
\begin{equation}
\Gamma_{\pi_{1\mu}^{hyb}\rightarrow b_{1}\pi}=2\frac{k_{1}}{8\pi m_{\pi
_{1}^{hyb}}^{2}}\left\{  \frac{1}{3}G_{0}^{2}{\lambda_{1}^{hyb}}^{2}Z_{\pi
}^{2}\left[  2+\frac{(m_{\pi_{1}^{hyb}}^{2}+m_{b_{1}}^{2}-m_{\pi}^{2})^{2}%
}{4m_{A}^{2}m_{B_{1}}^{2}}\right]  \right\}  \text{ .}%
\end{equation}
Similar expressions hold for all other possible decay widths described by the
first term. The results are listed in Table \ref{tabIII-IV}.

\subsection{Second term of the Lagrangian of Eq. (\ref{hybriddecay})}

The second term of the effective Lagrangian (\ref{hybriddecay})
\begin{equation}
\mathcal{L}_{eLSM,2}^{\text{ hybrid-linear}}=i\lambda_{2}^{hyb}\mathrm{Tr}%
([L_{\mu}^{hyb},L^{\mu}]\Phi\Phi^{\dagger}+[R_{\mu}^{hyb},R^{\mu}%
]\Phi^{\dagger}\Phi)\, \label{hybriddecay2}%
\end{equation}
generates two- and three-body decays for hybrid mesons into (axial-)vector
mesons and (pseudo)scalar mesons.

Let us first check the invariance under $P$ and $C.$ Under $P$ the first term
transforms into%
\begin{equation}
\mathrm{Tr}\left(  [L_{\mu}^{hyb},L^{\mu}]\Phi\Phi^{\dagger}\right)
\overset{P}{\rightarrow}\mathrm{Tr}\left(  [R^{hyb,\mu},R_{\mu}]\Phi^{\dag
}\Phi\right)
\end{equation}
therefore $P$ is conserved since the first term goes into the second. Under
$C,$ the first term transforms as:%
\begin{align}
&  \mathrm{Tr}\left(  [L_{\mu}^{hyb},L^{\mu}]\Phi\Phi^{\dagger}\right)
\overset{C}{\rightarrow}\mathrm{Tr}\left(  [R_{\mu}^{hyb,t},-R_{\mu}^{t}%
]\Phi^{\dag t}\Phi^{t}\right) \\
&  =\mathrm{Tr}\left(  (-R_{\mu}^{hyb,t}R_{\mu}^{t}+R_{\mu}^{t}R_{\mu}%
^{hyb,t})\left(  \Phi\Phi^{\dag}\right)  ^{t}\right)  =\mathrm{Tr}\left(
R_{\mu}^{hyb}R_{\mu}-R_{\mu}R_{\mu}^{hyb})^{t}\left(  \Phi\Phi^{\dag}\right)
^{t}\right) \\
&  =\mathrm{Tr}\left(  [R_{\mu}^{hyb},R_{\mu}]^{t}\left(  \Phi\Phi^{\dag
}\right)  ^{t}\right)  =\mathrm{Tr}\left(  \left(  \Phi\Phi^{\dag}\right)
[R_{\mu}^{hyb},R_{\mu}]\right)  =\mathrm{Tr}\left(  [R_{\mu}^{hyb},R_{\mu
}]\left(  \Phi\Phi^{\dag}\right)  \right)  .
\end{align}
Hence, \ $C$ is also conserved.

Last, we check that the matrix that the Lagrangian is Hermitian:
\begin{equation}
\left\{  i\mathrm{Tr}\left(  [L_{\mu}^{hyb},L^{\mu}]\Phi\Phi^{\dagger}\right)
\right\}  ^{\dagger}=-i\mathrm{Tr}\left(  \Phi\Phi^{\dagger}[L^{\mu},L_{\mu
}^{hyb}]\right)  =i\mathrm{Tr}\left(  [L_{\mu}^{hyb},L^{\mu}]\Phi\Phi
^{\dagger}\right)  \text{ .}%
\end{equation}

In terms of the nonets with defined $J^{PC}$ we get:%
\begin{align}
\mathcal{L}_{eLSM,2}^{\text{ hybrid-linear}}  &  =2i\lambda_{2}^{hyb}%
\mathrm{Tr}\left(  [\Pi_{\mu}^{hyb},V^{\mu}](P^{2}+S^{2})\right)
-2\lambda_{2}^{hyb}\mathrm{Tr}\left(  [\Pi_{\mu}^{hyb},A^{\mu}][P,S]\right)
-\\
&  2\lambda_{2}^{hyb}\mathrm{Tr}\left(  [B_{\mu}^{hyb},V^{\mu}][P,S]\right)
+2i\lambda_{2}^{hyb}\mathrm{Tr}\left(  [B_{\mu}^{hyb},A^{\mu}](P^{2}%
+S^{2})\right)  \text{ .}%
\end{align}
We then obtain the decays:%
\begin{equation}
\Pi_{\mu}^{hyb}\rightarrow VPP,\text{ }\Pi_{\mu}^{hyb}\rightarrow VSS,\text{
}\Pi_{\mu}^{hyb}\rightarrow A^{\mu}PS\text{ . }%
\end{equation}
The first is relevant, the second is suppressed, but the third can be relevant
due to the shift $A^{\mu}\rightarrow Zw\partial^{\mu}P$ and the condensation
of $S,$ since $\Pi_{\mu}^{hyb}\rightarrow PPS$ and $\Pi_{\mu}^{hyb}\rightarrow
PP$ follow. (Basically, out of $PP$ only $KK$ is possible, but is very much suppressed).

For the hybrid nonet $B_{\mu}^{hyb}$ one has:
\begin{equation}
B_{\mu}^{hyb}\rightarrow VPS,\text{ }B_{\mu}^{hyb}\rightarrow A^{\mu}PP,\text{
}B_{\mu}^{hyb}\rightarrow A^{\mu}SS\text{. }%
\end{equation}
Out of the first decay above, $B_{\mu}^{hyb}\rightarrow VP$ emerges upon
condensation of one field\ $S$ (but turns out to be suppressed), and out of
the second, $B_{\mu}^{hyb}\rightarrow PPP$ is realized when\ $A^{\mu}$ is
shifted, see the Appendix 1.

Next, for both terms we verify the invariance under $C,$ $P$ and $\dag$:
\begin{align}
\mathrm{Tr}\left(  [\Pi_{\mu}^{hyb},V^{\mu}](P^{2}+S^{2})\right)  \overset
{P}{\rightarrow}\mathrm{Tr}\left(  [\Pi^{hyb,\mu},V_{\mu}]((-1)^{2}P^{2}%
+S^{2})\right)   &  =\mathrm{Tr}\left(  [\Pi_{\mu}^{hyb},V^{\mu}](P^{2}%
+S^{2})\right)  \text{ ;}\\
\mathrm{Tr}\left(  [\Pi_{\mu}^{hyb},V^{\mu}](P^{2}+S^{2})\right)  \overset
{C}{\rightarrow}\mathrm{Tr}\left(  [\Pi_{\mu}^{hyb,t},-V^{\mu,t}%
](P^{2,t}+S^{2,t})\right)   &  =\mathrm{Tr}\left(  [\Pi_{\mu}^{hyb},V^{\mu
}](P^{2}+S^{2})\right)  \text{ ;}%
\end{align}

\begin{equation}
\left\{  2i\lambda_{2}^{hyb}\mathrm{Tr}\left(  [\Pi_{\mu}^{hyb},V^{\mu}%
](P^{2}+S^{2})\right)  \right\}  ^{\dag}\rightarrow-2i\lambda_{2}%
^{hyb}\mathrm{Tr}\left(  (P^{2}+S^{2})^{\dag}[\Pi_{\mu}^{hyb},V^{\mu}]^{\dag
}\right)  =2i\lambda_{2}^{hyb}\mathrm{Tr}\left(  [\Pi_{\mu}^{hyb},V^{\mu
}](P^{2}+S^{2})\right)  \text{ .}%
\end{equation}
The same transformations can be checked for the other terms. Next, we turn to
the two- and three-body decays described by this interaction term.

\subsubsection{\textbf{Two-body decay rates}}

As an example of a two-body decay, let us consider the case $b_{1\mu}%
^{hyb,0}\rightarrow a_{0}^{-}\pi^{+},$ which is described by the following
part of the Lagrangian:
\[
\mathcal{L}_{b_{1}a_{0}\pi}=\lambda_{2}^{hyb}\,Z_{\pi}w_{\pi}\,\phi_{N}\left[
b_{1\mu}^{0hyb}(a_{0}^{-}\partial^{\mu}\pi^{+}-a_{0}^{+}\partial^{\mu}\pi
^{-})+b_{1\mu}^{+hyb}(a_{0}^{0}\partial^{\mu}\pi^{-}-a_{0}^{-}\partial^{\mu
}\pi^{0})+b_{1\mu}^{-hyb}(a_{0}^{+}\partial^{\mu}\pi^{0}-a_{0}^{0}%
\partial^{\mu}\pi^{+})\right]
\]
We compute the decay width as
\begin{equation}
\Gamma_{b_{1}^{0hyb}\rightarrow a_{0}^{-}\pi^{+}}=\frac{|\overrightarrow
{k}_{1}|}{24\pi\,m_{b_{1}^{hyb}}^{2}}{\lambda_{2}^{hyb}}^{2}\left(  Z_{\pi
}w_{\pi}\right)  ^{2}\phi_{N}^{2}\,\left[  -m_{\pi}^{2}+\frac{(m_{b_{1}^{hyb}%
}^{2}+m_{\pi}^{2}-m_{a_{0}}^{2})^{2}}{4m_{b_{1}^{hyb}}^{2}}\right]  \,.
\end{equation}
The other decay channels of this type are calculated in a similar way and the
results are listed in (the left part of) Table \ref{tabV-VI}.

Next, we show why the decays $\pi_{1}^{hyb}\rightarrow\bar{K}K$, $\eta
_{1,N}^{hyb}\rightarrow\bar{K}K,$ and $\eta_{1,S}^{hyb}\rightarrow\bar{K}K$
vanish. For $\pi_{1}^{0,hyb}\rightarrow\bar{K}K,$ the corresponding Lagrangian is%

\begin{equation}
\mathcal{L}_{\pi_{1}^{0,hyb}\bar{K}K}=-\frac{1}{4}\lambda_{2}^{hyb}%
\,Z_{K}\,w_{K}\,(\phi_{N}-\sqrt{2}\phi_{S})\pi_{1}^{0hyb,\mu}\left[
\overline{K}^{0}\partial_{\mu}K^{0}+K^{0}\partial_{\mu}\overline{K}^{0}%
-K^{+}\partial_{\mu}K^{-}-K^{-}\partial_{\mu}K^{+}\right]  \text{ }.
\label{pi0hybkk}%
\end{equation}
This term follows from $-2\lambda_{2}^{hyb}Tr\left(  [\Pi_{\mu}^{hyb},A^{\mu
}][P,S]\right)  $ upon shifting $A^{\mu}$ according to Eq. (\ref{shifttot})
and setting $S=\Phi_{0}$ as in\ Eq. (\ref{cond}) (thus, flavor symmetry is
slightly broken). One can verify that this term is parity and $C$-invariant
thanks to the combination $\overline{K}^{0}\partial_{\mu}K^{0}+K^{0}%
\partial_{\mu}\overline{K}^{0}$. Moreover, isospin is also conserved, since
the quantity $\left[  \overline{K}^{0}\partial_{\mu}K^{0}+K^{0}\partial_{\mu
}\overline{K}^{0}-K^{+}\partial_{\mu}K^{-}-K^{-}\partial_{\mu}K^{+}\right]  $
has $I=1$ (intuitively, it is proportional to $\propto\bar{u}u-\bar{d}d),$ in
agreement with the fact that the $\pi_{1}^{0hyb,\mu}$ state has also isospin
$1.$ Yet, this interaction term does not lead to any decay $\pi_{1}^{0,hyb}$:
for instance, upon considering $\pi_{1}^{hyb}\rightarrow K^{+}K^{-}$ , one can
easily verify this point by considering that the decay amplitude is
proportional to
\begin{equation}
\varepsilon_{\mu}^{(a)}(p)\cdot\left(  k_{1}^{\mu}+k_{2}^{\mu}\right)
=\varepsilon_{\mu}^{(a)}(p)\cdot p^{\mu}=0\text{ ,}%
\end{equation}
where $p=(m_{\pi_{1},}0)$ is the four-momentum of $\pi_{1}^{hyb}$ and the
$k_{1}$ and $k_{2}$ are the four-momenta of the decaying particles. Note, the
same result can be also obtained by rewriting the relevant Lagrangian term
as:
\begin{align}
\pi_{1}^{0hyb,\mu}\left[  K^{+}\partial_{\mu}K^{-}+K^{-}\partial_{\mu}%
K^{+}\right]   &  =\pi_{1}^{0hyb,\mu}\partial_{\mu}\left[  K^{+}K^{-+}\right]
=\partial_{\mu}\left[  \pi_{1}^{0hyb,\mu}K^{+}K^{-+}\right]  -\left[
\partial_{\mu}\pi_{1}^{0hyb,\mu}\right]  K^{+}K^{-+}\nonumber\\
&  \equiv\left[  \partial_{\mu}\pi_{1}^{0hyb,\mu}\right]  K^{+}K^{-+}\text{,}%
\end{align}
where the full derivative can be neglected and the term proportional to
$\partial_{\mu}\pi_{1}^{0hyb,\mu}$ implies that no decay for $\pi_{1}^{hyb}$
is allowed as a consequence of the $\varepsilon_{\mu}^{(a)}(p)\cdot p^{\mu
}=0.$ Hence: $\Gamma_{\pi_{1}^{hyb,0}\rightarrow\bar{K}K}=0.$ Indeed, this
result is in agreement with $C$-parity, since this decay would violate it: the
initial state $\pi_{1}^{hyb,0}$ has by definition $C$-parity $+1,$ while the
final kaon-antikaon would necessarily have $C=-1.$ This can be seen by
considering the two-kaon state (schematically) as
\begin{equation}
\left\vert K^{+}(\mathbf{k}_{1})K^{-}(-\mathbf{k}_{1})\right\rangle \left\vert
L=1\right\rangle \text{ ,}%
\end{equation}
where $L=1$ assures that parity is $P=(-1)^{L}=-1$. Under $C$-transformation
this ket changes into%
\begin{equation}
\left\vert K^{-}(\mathbf{k}_{1})K^{+}(-\mathbf{k}_{1})\right\rangle \left\vert
L=1\right\rangle =-\left\vert K^{-}(\mathbf{k}_{1})K^{+}(-\mathbf{k}%
_{1})\right\rangle \left\vert L=1\right\rangle \text{ ,}%
\end{equation}
where in the last passage we took into account that the switching the momenta
is equivalent to a parity transformation. Hence, the $K^{+}K^{-}$ final state
has $C=-1.$

It is indeed interesting to notice that a Lagrangian term proportional to
$\left[  \partial_{\mu}\pi_{1}^{0hyb,\mu}\right]  K^{+}K^{-+}$ does not lead
to any decay of $\pi_{1}^{0hyb}$ and, in general, vanishes for any process in
which $\pi_{1}^{0hyb,\mu}$ is on-shell. Yet, the Lagrangian term cannot be set
to zero. In principle, it has a non-vanishing contribution to processes in
which $\pi_{1}^{0hyb,\mu}$ appear as a virtual particle. It has, for instance,
a very small but nonzero contribution to $K\bar{K}$ scattering.

Similarly, $\pi_{1}^{+hyb}\rightarrow K^{+}\bar{K}^{0}$ and $\pi_{1}%
^{-hyb}\rightarrow K^{-}K^{0}$ vanish since the interaction term have a form
analogous to Eq. (\ref{pi0hybkk}). This is in agreement with $G$-parity, where
$G=Ce^{i\pi I_{2}},$ $I_{2}$ being the second component of the isospin
operator $\vec{I}.$ The state $\pi_{1}^{+hyb}$ is not an eigenstate of $C$
parity, but is an eigenstate of $G$-parity with eigenvalue $-1.$ Yet, the
final state
\begin{equation}
\left\vert K^{+}(\mathbf{k}_{1})\bar{K}^{0}(-\mathbf{k}_{1})\right\rangle
\left\vert L=1\right\rangle
\end{equation}
has $G$ parity $+1,$ as one can verify by recalling that under $G$ the
transformations $K^{+}\rightarrow\bar{K}^{0}$ and $\bar{K}^{0}\rightarrow
-K^{+}$ hold. Quite interestingly, the decays $\pi_{1}^{+hyb}\rightarrow
K^{+}\bar{K}^{0}$ and $\pi_{1}^{-hyb}\rightarrow K^{-}K^{0}$ become possible
if isospin breaking is considered. Namely, the decay amplitudes would be
proportional to $\phi_{U}-\phi_{D},$ where $\phi_{U}$ and $\phi_{D}$
correspond to the $\bar{u}u$ and $\bar{d}d$ condensates, respectively (in the
present version of the model $\phi_{U}=\phi_{D}=\phi_{N}/\sqrt{2}$)$.$ The
inclusion of isospin violation of the eLSM is an interesting subject on its
own that can be investigated in the future.

Finally, we can summarize the result as
\begin{equation}
\Gamma_{\pi_{1}^{hyb}\rightarrow\bar{K}K}=0\text{.}%
\end{equation}

Next, the decay $\eta_{1,N}^{hyb}\rightarrow\bar{K}K$ also vanishes, since the
Lagrangian reads%
\[
\mathcal{L}_{\eta_{1,N}^{hyb}KK}=\frac{-1}{4}\lambda_{2}^{hyb}Z_{K}%
\,w_{K}\,(\phi_{N}-\sqrt{2}\phi_{S})\eta_{1,N}^{hyb,\mu}\left[  \overline
{K}^{0}\partial_{\mu}K^{0}+K^{0}\partial_{\mu}\overline{K}^{0}+K^{+}%
\partial_{\mu}K^{-}+K^{-}\partial_{\mu}K^{+}\right]  \,
\]
thus leading to the same discussion above. The same argument applies for
$\eta_{1,S}^{hyb}\rightarrow\bar{K}K.$ Both decays would violate $C$-parity.

\subsubsection{\textbf{Tree-body decay rates}}

We present the decay amplitudes for the three-body decay rates, which are
extracted from the Lagrangian (\ref{hybriddecay2}) and are used to compute for
the three-body decay widths. We use the following notations:
\begin{align*}
&  k_{1}\cdot k_{2}=\frac{m_{12}^{2}-m_{2}^{2}-m_{2}^{2}}{2}\,,\\
&  k\cdot k_{1}=m_{1}^{2}+\frac{m_{12}^{2}-m_{1}^{2}-m_{2}^{2}}{2}%
+\frac{m_{13}^{2}-m_{2}^{2}-m_{3}^{2}}{2}\,,\\
&  k\cdot k_{2}=k_{1}\cdot k_{2}+\frac{m_{12}^{2}-m_{1}^{2}-m_{2}^{2}}%
{2}+\frac{m_{23}^{2}-m_{2}^{2}-m_{3}^{2}}{2}\,,\\
&  m_{13}^{2}=M^{2}+m_{1}^{2}+m_{2}^{2}+m_{3}^{2}-m_{12}^{2}-m_{23}^{2}\,.
\end{align*}
\newline The decay amplitude for $\pi_{1}^{hyb}\rightarrow K^{\ast}K\pi$
channel\newline%
\begin{equation}
\left\vert -iM_{\pi_{1}^{0hyb}\rightarrow K^{\ast0}\overline{K}^{0}\pi^{0}%
}\right\vert =\frac{1}{16}\lambda_{2}^{hyb^{2}}\,Z_{K}^{2}\,Z_{\pi}^{2}%
.\frac{1}{9}\left[  2+\frac{(k\cdot k_{1})^{2}}{M^{2}m_{1}^{2}}\right]  \,,
\end{equation}
where the quantities $k,\,k_{1},\,k_{2},\,$and $k_{3}$ refer to the fields
$\pi_{1}^{0hyb},\,K^{\ast0},\,\overline{K}^{0},$ and $\pi^{0}$, respectively.
\newline

For instance, the decay width $b_{1}^{0hyb}\rightarrow\overline{K}^{0}K^{0}%
\pi^{0}$ reads
\begin{equation}
\Gamma_{b_{1}^{0hyb}\rightarrow\overline{K}^{0}K^{0}\pi^{0}}=\frac
{F_{b_{1}^{hyb}KK\pi}^{2}}{96\,(2\pi)^{3}\,M^{3}}\int_{(m_{1}+m_{2})^{2}%
}^{(M-m_{3})^{2}}\int_{(m_{23})_{\min}}^{(m_{23})_{\max}}\left[
\frac{\left\vert k\cdot k_{2}-k\cdot k_{1}\right\vert ^{2}}{M^{2}}-(m_{1}%
^{2}+m_{2}^{2}-2k_{1}\cdot k_{2})\right]  \,dm_{23}^{2}\,dm_{12}^{2}\>,
\end{equation}
and
\begin{equation}
F_{b_{1}^{hyb}KK\pi}=\frac{1}{4}\lambda_{2}^{hyb}Z_{\pi}\,Z_{K}\,\text{ ,}
\label{FD1Dpi}%
\end{equation}
Analogous expressions hold for the other channels and the results can be found
in the right part of Table \ref{tabV-VI} and the left part of Table
\ref{tabVII-VIII}.

\subsection{Third term of the Lagrangian of Eq. (\ref{hybriddecay})}

The third term of the effective Lagrangian (\ref{hybriddecay}) generate
two-body decays for hybrid mesons into (axial-)vector mesons and
(pseudo)scalar mesons, which are written as
\[
\mathcal{L}_{eLSM,3}^{\text{ hybrid-linear}}=\alpha^{hyb}\mathrm{Tr}(\tilde
{L}_{\mu\nu}^{hyb}\Phi R^{\mu\nu}\Phi^{\dagger}-\tilde{R}_{\mu\nu}^{hyb}%
\Phi^{\dagger}L^{\mu\nu}\Phi)\,.
\]
We first check the invariance under $P$ and $C.$ Parity is conserved because
the first term transforms into the second:
\begin{equation}
\mathrm{Tr}(\tilde{L}_{\mu\nu}^{hyb}\Phi R^{\mu\nu}\Phi^{\dagger})\overset
{P}{\rightarrow}\mathrm{Tr}((-\tilde{R}_{\mu\nu}^{hyb})\Phi^{\dagger}L^{\mu
\nu}\Phi)\text{.}%
\end{equation}
Note, the extra minus is due to the fact that the hybrid field is dual.

Under $C$ one has
\begin{equation}
\mathrm{Tr}(\tilde{L}_{\mu\nu}^{hyb}\Phi R^{\mu\nu}\Phi^{\dagger})\overset
{C}{\rightarrow}\mathrm{Tr}(\tilde{R}_{\mu\nu}^{hyb,t}\Phi^{t}(-R^{\mu\nu
,t})\Phi)\text{ ,}%
\end{equation}
hence $C$-invariance is preserved. As a last point, we check Hermiticity:%
\begin{equation}
\mathrm{Tr}(\tilde{L}_{\mu\nu}^{hyb}\Phi R^{\mu\nu}\Phi^{\dagger})^{\dag
}=\mathrm{Tr}(\Phi R^{\mu\nu\dag}\Phi^{\dagger}\tilde{L}_{\mu\nu}^{hyb\dag
})=\mathrm{Tr}(\Phi R^{\mu\nu}\Phi^{\dagger}\tilde{L}_{\mu\nu}^{hyb}%
)=\mathrm{Tr}(\tilde{L}_{\mu\nu}^{hyb}\Phi^{\dagger}R^{\mu\nu}\Phi)\text{ .}%
\end{equation}
In terms of the nonets, we isolate the following relevant terms relevant for
the two-body decays (obtained considering one condensation of the field $\Phi
$):
\begin{align}
\mathcal{L}_{eLSM,3}^{\text{ hybrid-linear}}  &  =2\alpha^{hyb}\mathrm{Tr}%
(\tilde{\Pi}_{\mu\nu}^{hyb}(-i\Phi_{0}V^{\mu\nu}P+iPV^{\mu\nu}\Phi
_{0})\,+\nonumber\\
&  2\alpha^{hyb}\mathrm{Tr}(\tilde{B}_{\mu\nu}^{hyb}(i\Phi_{0}A^{\mu\nu
}P-iPA^{\mu\nu}\Phi_{0})\,+...
\end{align}
Then, we obtain decays of the type $\tilde{\Pi}^{hyb}\rightarrow PV$ and
$\tilde{B}_{\mu\nu}^{hyb}\rightarrow AP.$ Using Eq. (\ref{apprphi}), the
flavor-invariant piece is given by
\begin{equation}
\mathcal{L}_{eLSM,3}^{\text{ hybrid-linear}}=2i\alpha^{hyb}\frac{\phi_{N}}%
{2}\left\{  \mathrm{Tr}(\tilde{\Pi}_{\mu\nu}^{hyb}[P,V^{\mu\nu}%
])\,-\mathrm{Tr}(\tilde{B}_{\mu\nu}^{hyb}([P,A^{\mu\nu}])\,\right\}
+...\nonumber
\end{equation}

For completeness, we verify the correctness of the first term upon checking
the invariance under $C,$ $P,$ and $\dag$:%
\begin{align}
\mathrm{Tr}(\tilde{\Pi}_{\mu\nu}^{hyb}[P,V^{\mu\nu}])\overset{P}{\rightarrow
}\mathrm{Tr}(-\tilde{\Pi}^{hyb,\mu\nu}[-P,V_{\mu\nu}])  &  =\mathrm{Tr}%
(\tilde{\Pi}_{\mu\nu}^{hyb}[P,V^{\mu\nu}])\text{ ;}\\
\mathrm{Tr}(\tilde{\Pi}_{\mu\nu}^{hyb}[P,V^{\mu\nu}])\overset{C}{\rightarrow
}\mathrm{Tr}(\tilde{\Pi}_{\mu\nu}^{hyb,t}[P^{t},-V_{\mu\nu}^{t}])  &
=\mathrm{Tr}(\tilde{\Pi}_{\mu\nu}^{hyb}[P,V^{\mu\nu}])\text{ ;}%
\end{align}%
\begin{equation}
\left\{  i\mathrm{Tr}(\tilde{\Pi}_{\mu\nu}^{hyb}[P,V^{\mu\nu}])\,\right\}
^{\dag}=-i\mathrm{Tr}([P,V^{\mu\nu}]^{\dag}\tilde{\Pi}_{\mu\nu}^{hyb,\dag
})\,=i\mathrm{Tr}(\tilde{\Pi}_{\mu\nu}^{hyb}[P,V^{\mu\nu}])\text{ .}%
\end{equation}

As an example, let us consider the following explicit term in the Lagrangian:
\begin{equation}
\mathcal{L}_{eLSM-\alpha^{hyb}}^{\text{ hybrid-linear}}=2i\alpha^{hyb}\phi
_{N}Z_{\pi}\tilde{\pi}_{1\,\mu\nu}^{+hyb}(\pi^{-}\rho^{0,\mu\nu}+\pi^{0}%
\rho^{-,\mu\nu})+...\text{ ,}%
\end{equation}
for which the explicit decay width reads:%
\begin{equation}
\Gamma_{\pi_{1\mu}^{hyb}\rightarrow\rho\pi}=2\frac{k_{1}}{8\pi m_{\pi
_{1}^{hyb}}^{2}}\left[  \left(  2\alpha^{hyb}\phi_{N}Z_{\pi}\right)  ^{2}%
\frac{8}{3}m_{\pi_{1}^{hyb}}^{2}k_{1}^{2}\right]
\end{equation}
Similar decay widths hold for the other channels. The corresponding results
can be found in the right part of Table \ref{tabVII-VIII}.

\subsection{Fourth term of the Lagrangian of Eq. (\ref{hybriddecay}): the
anomaly term}

\label{appC4}

Finally, we consider the fourth (and the last) term of Eq. (\ref{hybriddecay}%
):%
\begin{equation}
\mathcal{L}_{eLSM,4}^{\text{ hybrid-linear}}=\beta_{A}^{hyb}(\det\Phi-\det
\Phi^{\dag})\mathrm{Tr}(L_{\mu}^{hyb}(\partial^{\mu}\Phi\cdot\Phi^{\dag}%
-\Phi\cdot\partial^{\mu}\Phi^{\dag})-R_{\mu}^{hyb}(\partial^{\mu}\Phi^{\dag
}\cdot\Phi-\Phi^{\dag}\cdot\partial^{\mu}\Phi))\text{ .}%
\end{equation}
As for the other cases, we check the transformation properties. Under parity,
the first term transforms as:%
\begin{align}
&  (\det\Phi-\det\Phi^{\dag})\mathrm{Tr}(L_{\mu}^{hyb}(\partial^{\mu}\Phi
\cdot\Phi^{\dag}))\overset{P}{\rightarrow}(\det\Phi^{\dag}-\det\Phi
)\mathrm{Tr}(R_{\mu}^{hyb}(\partial^{\mu}\Phi^{\dag}\cdot\Phi))\nonumber\\
&  =\beta_{A}^{hyb}(\det\Phi-\det\Phi^{\dag})\mathrm{Tr}(-R_{\mu}%
^{hyb}(\partial^{\mu}\Phi^{\dag}\cdot\Phi))\text{ ,}%
\end{align}
therefore the first term transforms into the third. Similarly, the second
converts into the fourth, assuring that $P$ is preserved.

Next, one has%
\begin{align}
&  (\det\Phi-\det\Phi^{\dag})\mathrm{Tr}(L_{\mu}^{hyb}(\partial^{\mu}\Phi
\cdot\Phi^{\dag}))\overset{C}{\rightarrow}(\det\Phi^{\dag,t}-\det\Phi
^{t})\mathrm{Tr}(R_{\mu}^{hyb,t}(\partial^{\mu}\Phi^{t}\cdot\Phi^{\dag
t}))\nonumber\\
&  =(\det\Phi-\det\Phi^{\dag})\mathrm{Tr}(R_{\mu}^{hyb,t}(\Phi^{\dag}%
\partial^{\mu}\Phi)^{t})=(\det\Phi-\det\Phi^{\dag})\mathrm{Tr}(\Phi^{\dag
}\partial^{\mu}\Phi R_{\mu}^{hyb})\nonumber\\
&  =(\det\Phi-\det\Phi^{\dag})\mathrm{Tr}(R_{\mu}^{hyb}\Phi^{\dag}%
\cdot\partial^{\mu}\Phi)\text{ ,}%
\end{align}
which shows that the first term converts into the fourth. Similarly, the
second goes into the third.

Finally, we show that the Lagrangian is Hermitian.\ For the first term:
\begin{align}
\left\{  (\det\Phi-\det\Phi^{\dag})\mathrm{Tr}(L_{\mu}^{hyb}(\partial^{\mu
}\Phi\cdot\Phi^{\dag}))\right\}  ^{\dag}  &  =(\det\Phi^{\dag}-\det
\Phi)\mathrm{Tr}(\Phi\partial^{\mu}\Phi^{\dag}L_{\mu}^{hyb})\nonumber\\
&  =-(\det\Phi-\det\Phi^{\dag})\mathrm{Tr}(L_{\mu}^{hyb}\Phi\partial^{\mu}%
\Phi^{\dag})\text{ ,}%
\end{align}
showing that the first converts into the second.

In terms of the fields, we recall that \cite{Olbrich:2017fsd}
\begin{equation}
\det\Phi-\det\Phi^{\dag}=i\frac{Z_{\pi}}{2}\sqrt{\frac{3}{2}}\phi_{N}^{2}%
\eta_{0}+...\text{,}%
\end{equation}
where dots refer to flavor breaking corrections and to terms involving two or
more fields. Then, upon condensation of one $\Phi$ and using Eq.
(\ref{apprphi}):
\begin{align}
\mathcal{L}_{eLSM,4}^{\text{ hybrid-linear}}  &  =i\beta_{A}^{hyb}\frac
{Z_{\pi}}{4}\sqrt{\frac{3}{2}}\phi_{N}^{3}\eta_{0}\mathrm{Tr}(L_{\mu}%
^{hyb}(\partial^{\mu}\Phi-\partial^{\mu}\Phi^{\dag})-R_{\mu}^{hyb}%
(\partial^{\mu}\Phi^{\dag}-\partial^{\mu}\Phi))+...\\
&  =i\beta_{A}^{hyb}\frac{Z_{\pi}}{4}\sqrt{\frac{3}{2}}\phi_{N}^{3}\eta
_{0}\mathrm{Tr}(L_{\mu}^{hyb}(2i\partial^{\mu}P)-R_{\mu}^{hyb}(-2i\partial
^{\mu}P))+...\\
&  =-\beta_{A}^{hyb}Z_{\pi}\sqrt{\frac{3}{2}}\phi_{N}^{3}\eta_{0}%
\mathrm{Tr}(\Pi_{\mu}^{hyb}\partial^{\mu}P)+...
\end{align}
which described the decay $\Pi_{\mu}^{hyb}\mapsto P\eta_{0}$ and hence
$\Pi_{\mu}^{hyb}\mapsto P\eta$ and $\Pi_{\mu}^{hyb}\mapsto P\eta^{\prime}.$

As an example, let us report the explicit form of decay width of the process
$\pi_{1}\rightarrow\pi\eta$:%
\begin{equation}
\Gamma_{\pi_{1}\rightarrow\pi\eta}=\frac{k_{1}}{8\pi m_{\pi_{1}}^{2}}\left(
\beta_{A}^{hyb}Z_{\pi}^{2}\frac{\sqrt{3}}{2}\phi_{N}^{3}\right)  ^{2}\frac
{1}{3}\left[  -m_{\pi}^{2}+\frac{\left(  m_{\pi_{1}^{hyb}}^{2}+m_{\pi}%
^{2}-m_{\eta}^{2}\right)  ^{2}}{m_{\pi_{1}^{hyb}}^{2}}\right]  \text{ .}%
\end{equation}
The other decays listed in Table \ref{tabIX} are calculated following the same steps.

Note, the decays $\eta_{1,N}\rightarrow\eta\eta$ and $\eta_{1,S}%
\rightarrow\eta\eta$ do not take place because the final state consists of
identical particles. The total amplitude is the sum of two terms -the direct
and the crossed Feynman diagrams- which cancel each other. More precisely, the
Lagrangian above contains a terms proportional to
\begin{equation}
\eta_{1,N}^{\mu}\left(  \partial_{\mu}\eta\right)  \eta\text{ ,}%
\end{equation}
out of which the full amplitude is proportional to $A^{\alpha}\propto
\varepsilon^{(\alpha)\mu}(p=0)(k_{1,\mu}+k_{2,\mu})=\varepsilon^{(\alpha)\mu
}(0)p_{\mu}=0.$ (For non-identical particles the amplitude is proportional to
$\varepsilon^{(\alpha)\mu}(p=0)k_{1,\mu},$ which does not vanish.) One can
reach the same conclusion by rewriting the Lagrangian term as%
\begin{equation}
\eta_{1,N}^{\mu}\left(  \partial_{\mu}\eta\right)  \eta=\eta_{1,N}^{\mu
}\left(  \partial_{\mu}\eta^{2}/2\right)  =\partial^{\mu}\left(  \eta
_{1,N}^{\mu}\eta^{2}/2\right)  -(\partial^{\mu}\eta_{1,N}^{\mu})\eta
^{2}/2\text{ ,}%
\end{equation}
which does not lead to any decay, since the first term on the r.h.s. is a full
derivative and the second term contains $(\partial^{\mu}\eta_{1,N}^{\mu}),$
which vanishes as a consequence of the Proca equation.

\bibliographystyle{unsrt}
\bibliography{hybrid}

\begin{thebibliography}{10}

\bibitem{Meyer:2015eta}
C.~A. Meyer and E.~S. Swanson.
\newblock {Hybrid Mesons}.
\newblock {\em Prog. Part. Nucl. Phys.}, 82:21--58, 2015.

\bibitem{Lebed:2016hpi}
Richard~F. Lebed, Ryan~E. Mitchell, and Eric~S. Swanson.
\newblock {Heavy-Quark QCD Exotica}.
\newblock {\em Prog. Part. Nucl. Phys.}, 93:143--194, 2017.

\bibitem{Michael:1999ge}
Christopher Michael.
\newblock {Quarkonia and hybrids from the lattice}.
\newblock page hf8/001, 1999.
\newblock [PoShf8,001(1999)].

\bibitem{Dudek:2009qf}
Jozef~J. Dudek, Robert~G. Edwards, Michael~J. Peardon, David~G. Richards, and
  Christopher~E. Thomas.
\newblock {Highly excited and exotic meson spectrum from dynamical lattice
  QCD}.
\newblock {\em Phys. Rev. Lett.}, 103:262001, 2009.

\bibitem{Dudek:2010wm}
Jozef~J. Dudek, Robert~G. Edwards, Michael~J. Peardon, David~G. Richards, and
  Christopher~E. Thomas.
\newblock {Toward the excited meson spectrum of dynamical QCD}.
\newblock {\em Phys. Rev.}, D82:034508, 2010.

\bibitem{Dudek:2013yja}
Jozef~J. Dudek, Robert~G. Edwards, Peng Guo, and Christopher~E. Thomas.
\newblock {Toward the excited isoscalar meson spectrum from lattice QCD}.
\newblock {\em Phys. Rev.}, D88(9):094505, 2013.

\bibitem{Liu:2012ze}
Liuming Liu, Graham Moir, Michael Peardon, Sinead~M. Ryan, Christopher~E.
  Thomas, Pol Vilaseca, Jozef~J. Dudek, Robert~G. Edwards, Balint Joo, and
  David~G. Richards.
\newblock {Excited and exotic charmonium spectroscopy from lattice QCD}.
\newblock {\em JHEP}, 07:126, 2012.

\bibitem{Moir:2013ub}
Graham Moir, Michael Peardon, Sinead~M. Ryan, Christopher~E. Thomas, and
  Liuming Liu.
\newblock {Excited spectroscopy of charmed mesons from lattice QCD}.
\newblock {\em JHEP}, 05:021, 2013.

\bibitem{Tanabashi:2018oca}
M.~Tanabashi et~al.
\newblock {Review of Particle Physics}.
\newblock {\em Phys. Rev.}, D98(3):030001, 2018.

\bibitem{Akhunzyanov:2018lqa}
M.~Aghasyan et~al.
\newblock {Light isovector resonances in $\pi^- p \to \pi^-\pi^-\pi^+ p$ at 190
  GeV/${\it c}$}.
\newblock {\em Phys. Rev.}, D98(9):092003, 2018.

\bibitem{Dobbs:2017vjw}
Sean Dobbs.
\newblock {Searching for Hybrid Mesons with GlueX}.
\newblock {\em PoS}, Hadron2017:047, 2018.

\bibitem{Rizzo:2016qvl}
A.~Rizzo.
\newblock {The hybrid mesons quest: the MesonEx experiment at Jefferson
  Laboratory}.
\newblock {\em J. Phys. Conf. Ser.}, 689(1):012022, 2016.

\bibitem{Ablikim:2005um}
M.~Ablikim et~al.
\newblock {Observation of a resonance X(1835) in J / psi -\&gt; gamma pi+ pi-
  eta-prime}.
\newblock {\em Phys. Rev. Lett.}, 95:262001, 2005.

\bibitem{Kochelev:2005vd}
Nikolai Kochelev and Dong-Pil Min.
\newblock {X(1835) as the lowest mass pseudoscalar glueball and proton spin
  problem}.
\newblock {\em Phys. Lett.}, B633:283--288, 2006.

\bibitem{Ablikim:2010au}
M.~Ablikim et~al.
\newblock {Confirmation of the $X(1835)$ and observation of the resonances
  $X(2120)$ and $X(2370)$ in $J/\psi\to \gamma \pi^+\pi^-\eta^\prime$}.
\newblock {\em Phys. Rev. Lett.}, 106:072002, 2011.

\bibitem{Lutz:2009ff}
M.~F.~M. Lutz et~al.
\newblock {Physics Performance Report for PANDA: Strong Interaction Studies
  with Antiprotons}.
\newblock 2009.

\bibitem{Parganlija:2012fy}
Denis Parganlija, Peter Kovacs, Gyorgy Wolf, Francesco Giacosa, and Dirk~H.
  Rischke.
\newblock {Meson vacuum phenomenology in a three-flavor linear sigma model with
  (axial-)vector mesons}.
\newblock {\em Phys. Rev.}, D87(1):014011, 2013.

\bibitem{Janowski:2014ppa}
Stanislaus Janowski, Francesco Giacosa, and Dirk~H. Rischke.
\newblock {Is f0(1710) a glueball?}
\newblock {\em Phys. Rev.}, D90(11):114005, 2014.

\bibitem{Parganlija:2010fz}
Denis Parganlija, Francesco Giacosa, and Dirk~H. Rischke.
\newblock {Vacuum Properties of Mesons in a Linear Sigma Model with Vector
  Mesons and Global Chiral Invariance}.
\newblock {\em Phys. Rev.}, D82:054024, 2010.

\bibitem{Parganlija:2016yxq}
Denis Parganlija and Francesco Giacosa.
\newblock {Excited Scalar and Pseudoscalar Mesons in the Extended Linear Sigma
  Model}.
\newblock {\em Eur. Phys. J.}, C77(7):450, 2017.

\bibitem{Eshraim:2012jv}
Walaa~I. Eshraim, Stanislaus Janowski, Francesco Giacosa, and Dirk~H. Rischke.
\newblock {Decay of the pseudoscalar glueball into scalar and pseudoscalar
  mesons}.
\newblock {\em Phys. Rev.}, D87(5):054036, 2013.

\bibitem{Eshraim:2016mds}
Walaa~I. Eshraim and Stefan Schramm.
\newblock {Decay modes of the excited pseudoscalar glueball}.
\newblock {\em Phys. Rev.}, D95(1):014028, 2017.

\bibitem{Eshraim:2019sgr}
Walaa~I. Eshraim.
\newblock {Decay of the pseudoscalar glueball and its first excited state into
  scalar and pseudoscalar mesons and their first excited states }.
\newblock {\em Phys. Rev.}

\bibitem{Giacosa:2016hrm}
Francesco Giacosa, Julia Sammet, and Stanislaus Janowski.
\newblock {Decays of the vector glueball}.
\newblock {\em Phys. Rev.}, D95(11):114004, 2017.

\bibitem{Divotgey:2016pst}
Florian Divotgey, Peter Kovacs, Francesco Giacosa, and Dirk~H. Rischke.
\newblock {Low-energy limit of the extended Linear Sigma Model}.
\newblock {\em Eur. Phys. J.}, A54(1):5, 2018.

\bibitem{Eshraim:2014eka}
Walaa~I. Eshraim, Francesco Giacosa, and Dirk~H. Rischke.
\newblock {Phenomenology of charmed mesons in the extended Linear Sigma Model}.
\newblock {\em Eur. Phys. J.}, A51(9):112, 2015.

\bibitem{Eshraim:2018iea}
Walaa~I. Eshraim and Christian~S. Fischer.
\newblock {Hadronic decays of the (pseudo-)scalar charmonium states $\eta_{c}$
  and $\chi_{c0}$ in the extended Linear Sigma Model}.
\newblock {\em Eur. Phys. J.}, A54(8):139, 2018.

\bibitem{Gallas:2009qp}
Susanna Gallas, Francesco Giacosa, and Dirk~H. Rischke.
\newblock {Vacuum phenomenology of the chiral partner of the nucleon in a
  linear sigma model with vector mesons}.
\newblock {\em Phys. Rev.}, D82:014004, 2010.

\bibitem{Olbrich:2015gln}
Lisa Olbrich, Miklos Zetenyi, Francesco Giacosa, and Dirk~H. Rischke.
\newblock {Three-flavor chiral effective model with four baryonic multiplets
  within the mirror assignment}.
\newblock {\em Phys. Rev.}, D93(3):034021, 2016.

\bibitem{Olbrich:2017fsd}
Lisa Olbrich, Miklis Zetenyi, Francesco Giacosa, and Dirk~H. Rischke.
\newblock {Influence of the axial anomaly on the decay $N(1535) \rightarrow
  N\eta $}.
\newblock {\em Phys. Rev.}, D97(1):014007, 2018.

\bibitem{Lakaschus:2018rki}
Phillip Lakaschus, Justin~L.P. Mauldin, Francesco Giacosa, and Dirk~H. Rischke.
\newblock {Role of a four-quark and a glueball state in pion-pion and
  pion-nucleon scattering}.
\newblock {\em Phys. Rev. C}, 99(4):045203, 2019.

\bibitem{Kovacs:2016juc}
Peter Kovacs, Zsolt Szep, and Gyorgy Wolf.
\newblock {Existence of the critical endpoint in the vector meson extended
  linear sigma model}.
\newblock {\em Phys. Rev.}, D93(11):114014, 2016.

\bibitem{Tawfik:2014gga}
Abdel~Nasser Tawfik and Abdel~Magied Diab.
\newblock {Polyakov SU(3) extended linear- $\sigma$ model: Sixteen mesonic
  states in chiral phase structure}.
\newblock {\em Phys. Rev.}, C91(1):015204, 2015.

\bibitem{Gallas:2011qp}
Susanna Gallas, Francesco Giacosa, and Giuseppe Pagliara.
\newblock {Nuclear matter within a dilatation-invariant parity doublet model:
  the role of the tetraquark at nonzero density}.
\newblock {\em Nucl. Phys. A}, 872:13--24, 2011.

\bibitem{Heinz:2013hza}
Achim Heinz, Francesco Giacosa, and Dirk~H. Rischke.
\newblock {Chiral density wave in nuclear matter}.
\newblock {\em Nucl. Phys. A}, 933:34--42, 2015.

\bibitem{Thomas:2001kw}
Anthony~William Thomas and Wolfram Weise.
\newblock {\em {The Structure of the Nucleon}}.
\newblock Wiley, Germany, 2001.

\bibitem{Migdal:1982jp}
Alexander~A. Migdal and Mikhail~A. Shifman.
\newblock {Dilaton Effective Lagrangian in Gluodynamics}.
\newblock {\em Phys. Lett.}, 114B:445--449, 1982.

\bibitem{Gomm:1985ut}
R.~Gomm, P.~Jain, R.~Johnson, and J.~Schechter.
\newblock {Scale Anomaly and the Scalars}.
\newblock {\em Phys. Rev.}, D33:801, 1986.

\bibitem{tHooft:1986ooh}
Gerard 't~Hooft.
\newblock {How Instantons Solve the U(1) Problem}.
\newblock {\em Phys. Rept.}, 142:357--387, 1986.

\bibitem{Gomm:1984at}
H.~Gomm, O.~Kaymakcalan, and J.~Schechter.
\newblock {Anomalous Spin 1 Meson Decays From the Gauged {Wess-Zumino} Term}.
\newblock {\em Phys. Rev. D}, 30:2345, 1984.

\bibitem{Rodas:2018owy}
A.~Rodas et~al.
\newblock {Determination of the pole position of the lightest hybrid meson
  candidate}.
\newblock {\em Phys. Rev. Lett.}, 122(4):042002, 2019.

\bibitem{AmelinoCamelia:2010me}
G.~Amelino-Camelia et~al.
\newblock {Physics with the KLOE-2 experiment at the upgraded DA$\phi$NE}.
\newblock {\em Eur. Phys. J.}, C68:619--681, 2010.

\bibitem{Gui:2012gx}
Long-Cheng Gui, Ying Chen, Gang Li, Chuan Liu, Yu-Bin Liu, Jian-Ping Ma, Yi-Bo
  Yang, and Jian-Bo Zhang.
\newblock {Scalar Glueball in Radiative $J/\psi$ Decay on the Lattice}.
\newblock {\em Phys. Rev. Lett.}, 110(2):021601, 2013.

\bibitem{Brunner:2015oqa}
Frederic Bruenner, Denis Parganlija, and Anton Rebhan.
\newblock {Glueball Decay Rates in the Witten-Sakai-Sugimoto Model}.
\newblock {\em Phys. Rev.}, D91(10):106002, 2015.
\newblock [Erratum: Phys. Rev.D93,no.10,109903(2016)].

\bibitem{Brunner:2015yha}
Frederic Bruenner and Anton Rebhan.
\newblock {Nonchiral enhancement of scalar glueball decay in the
  Witten-Sakai-Sugimoto model}.
\newblock {\em Phys. Rev. Lett.}, 115(13):131601, 2015.

\bibitem{Brunner:2015oga}
Frederic Bruenner and Anton Rebhan.
\newblock {Constraints on the $\eta \eta'$ decay rate of a scalar glueball from
  gauge/gravity duality}.
\newblock {\em Phys. Rev.}, D92(12):121902, 2015.

\bibitem{Jaffe:2004ph}
R.~L. Jaffe.
\newblock {Exotica}.
\newblock {\em Phys. Rept.}, 409:1--45, 2005.

\bibitem{Pelaez:2003dy}
J.~R. Pelaez.
\newblock {On the Nature of light scalar mesons from their large N(c)
  behavior}.
\newblock {\em Phys. Rev. Lett.}, 92:102001, 2004.

\bibitem{Napsuciale:2004au}
Mauro Napsuciale and Simon Rodriguez.
\newblock {A Chiral model for anti-q q and anti-qq qq mesons}.
\newblock {\em Phys. Rev.}, D70:094043, 2004.

\bibitem{Fariborz:2003uj}
Amir~H. Fariborz.
\newblock {Isosinglet scalar mesons below 2-GeV and the scalar glueball mass}.
\newblock {\em Int. J. Mod. Phys.}, A19:2095--2112, 2004.

\bibitem{Fariborz:2005gm}
Amir~H. Fariborz, Renata Jora, and Joseph Schechter.
\newblock {Toy model for two chiral nonets}.
\newblock {\em Phys. Rev.}, D72:034001, 2005.

\bibitem{Maiani:2004uc}
L.~Maiani, F.~Piccinini, A.~D. Polosa, and V.~Riquer.
\newblock {A New look at scalar mesons}.
\newblock {\em Phys. Rev. Lett.}, 93:212002, 2004.

\bibitem{Giacosa:2006rg}
Francesco Giacosa.
\newblock {Strong and electromagnetic decays of the light scalar mesons
  interpreted as tetraquark states}.
\newblock {\em Phys. Rev.}, D74:014028, 2006.

\bibitem{Giacosa:2009qh}
Francesco Giacosa and Giuseppe Pagliara.
\newblock {Decay of light scalar mesons into vector-photon and into
  pseudoscalar mesons}.
\newblock {\em Nucl. Phys.}, A833:138--155, 2010.

\bibitem{Heupel:2012ua}
Walter Heupel, Gernot Eichmann, and Christian~S. Fischer.
\newblock {Tetraquark Bound States in a Bethe-Salpeter Approach}.
\newblock {\em Phys. Lett.}, B718:545--549, 2012.

\bibitem{Eichmann:2015cra}
Gernot Eichmann, Christian~S. Fischer, and Walter Heupel.
\newblock {The light scalar mesons as tetraquarks}.
\newblock {\em Phys. Lett.}, B753:282--287, 2016.

\bibitem{Pelaez:2015qba}
J.~R. Pelaez.
\newblock {From controversy to precision on the sigma meson: a review on the
  status of the non-ordinary $f_0(500)$ resonance}.
\newblock {\em Phys. Rept.}, 658:1, 2016.

\bibitem{vanBeveren:1986ea}
E.~van Beveren, T.~A. Rijken, K.~Metzger, C.~Dullemond, G.~Rupp, and J.~E.
  Ribeiro.
\newblock {A Low Lying Scalar Meson Nonet in a Unitarized Meson Model}.
\newblock {\em Z. Phys.}, C30:615--620, 1986.

\bibitem{Oller:1997ti}
J.~A. Oller and E.~Oset.
\newblock {Chiral symmetry amplitudes in the S wave isoscalar and isovector
  channels and the $\sigma$, f$_0$(980), a$_0$(980) scalar mesons}.
\newblock {\em Nucl. Phys.}, A620:438--456, 1997.
\newblock [Erratum: Nucl. Phys.A652,407(1999)].

\bibitem{Oller:1998hw}
J.~A. Oller, E.~Oset, and J.~R. Pelaez.
\newblock {Meson meson interaction in a nonperturbative chiral approach}.
\newblock {\em Phys. Rev.}, D59:074001, 1999.
\newblock [Erratum: Phys. Rev.D75,099903(2007)].

\bibitem{vanBeveren:2006ua}
Eef van Beveren, D.~V. Bugg, F.~Kleefeld, and G.~Rupp.
\newblock {The Nature of sigma, kappa, a(0)(980) and f(0)(980)}.
\newblock {\em Phys. Lett.}, B641:265--271, 2006.

\bibitem{Parganlija:2012xj}
Denis Parganlija.
\newblock {\em {Quarkonium Phenomenology in Vacuum}}.
\newblock PhD thesis, Frankfurt U., 2011.

\bibitem{Divotgey:2013jba}
Florian Divotgey, Lisa Olbrich, and Francesco Giacosa.
\newblock {Phenomenology of axial-vector and pseudovector mesons: decays and
  mixing in the kaonic sector}.
\newblock {\em Eur. Phys. J.}, A49:135, 2013.

\bibitem{Piotrowska:2017rgt}
Milena Piotrowska, Christian Reisinger, and Francesco Giacosa.
\newblock {Strong and radiative decays of excited vector mesons and predictions
  for a new $\phi(1930)$ resonance}.
\newblock {\em Phys. Rev.}, D96(5):054033, 2017.

\bibitem{Giacosa:2017pos}
Francesco Giacosa, Adrian Koenigstein, and Robert~D. Pisarski.
\newblock {How the axial anomaly controls flavor mixing among mesons}.
\newblock {\em Phys. Rev.}, D97(9):091901, 2018.

\bibitem{Giacosa:2007bn}
Francesco Giacosa and Giuseppe Pagliara.
\newblock {On the spectral functions of scalar mesons}.
\newblock {\em Phys. Rev. C}, 76:065204, 2007.

\bibitem{Schneitzer:2014rsa}
Jonas Schneitzer, Thomas Wolkanowski, and Francesco Giacosa.
\newblock {The role of the next-to-leading order triangle-shaped diagram in
  two-body hadronic decays}.
\newblock {\em Nucl. Phys. B}, 888:287--299, 2014.

\bibitem{Wolkanowski:2015lsa}
Thomas Wolkanowski, Francesco Giacosa, and Dirk~H. Rischke.
\newblock {$a_{0}(980)$ revisited}.
\newblock {\em Phys. Rev.}, D93(1):014002, 2016.

\bibitem{Boglione:2002vv}
M.~Boglione and M.~R. Pennington.
\newblock {Dynamical generation of scalar mesons}.
\newblock {\em Phys. Rev.}, D65:114010, 2002.

\bibitem{Wolkanowski:2015jtc}
Thomas Wolkanowski, M.~So?tysiak, and F.~Giacosa.
\newblock {$K_{0}^{\ast}(800)$ as a companion pole of $K_{0}^{\ast}(1430)$}.
\newblock {\em Nucl. Phys.}, B909:418--428, 2016.

\bibitem{Black:2000qq}
Deirdre Black, Amir~H. Fariborz, Sherif Moussa, Salah Nasri, and Joseph
  Schechter.
\newblock {Unitarized pseudoscalar meson scattering amplitudes in three flavor
  linear sigma models}.
\newblock {\em Phys. Rev. D}, 64:014031, 2001.

\bibitem{Chen:2005mg}
Y.~Chen et~al.
\newblock {Glueball spectrum and matrix elements on anisotropic lattices}.
\newblock {\em Phys. Rev.}, D73:014516, 2006.

\bibitem{Close:1994hc}
Frank~E. Close and Philip~R. Page.
\newblock {The Production and decay of hybrid mesons by flux tube breaking}.
\newblock {\em Nucl. Phys. B}, 443:233--254, 1995.

\bibitem{McNeile:2006bz}
C.~McNeile and Christopher Michael.
\newblock {Decay width of light quark hybrid meson from the lattice}.
\newblock {\em Phys. Rev. D}, 73:074506, 2006.

\bibitem{Bass:2001zs}
Steven~D. Bass and Eugenio Marco.
\newblock {Final state interaction and a light mass 'exotic' resonance}.
\newblock {\em Phys. Rev. D}, 65:057503, 2002.

\bibitem{Szczepaniak:2003vg}
Adam~P. Szczepaniak, Maciej Swat, Alex~R. Dzierba, and Scott Teige.
\newblock {Study of the eta pi and eta-prime pi spectra and interpretation of
  possible exotic J**PC = 1-+ mesons}.
\newblock {\em Phys. Rev. Lett.}, 91:092002, 2003.

\bibitem{Eshraim:2012rb}
Walaa~I. Eshraim, Stanislaus Janowski, Antje Peters, Klaus Neuschwander, and
  Francesco Giacosa.
\newblock {Interaction of the pseudoscalar glueball with (pseudo)scalar mesons
  and nucleons}.
\newblock {\em Acta Phys. Polon. Supp.}, 5:1101--1108, 2012.

\bibitem{Tornqvist:1995kr}
Nils~A. Tornqvist.
\newblock {Understanding the scalar meson q anti-q nonet}.
\newblock {\em Z. Phys.}, C68:647--660, 1995.

\bibitem{Tornqvist:1995ay}
Nils~A. Tornqvist and Matts Roos.
\newblock {Resurrection of the sigma meson}.
\newblock {\em Phys. Rev. Lett.}, 76:1575--1578, 1996.

\bibitem{Boglione:1997aw}
Mariaelena Boglione and M.~R. Pennington.
\newblock {Unquenching the scalar glueball}.
\newblock {\em Phys. Rev. Lett.}, 79:1998--2001, 1997.

\end{thebibliography}

\end{document}